\documentclass[aps,twocolumn,showpacs,amsmath,amssymb,pra,superscriptaddress,floatfix,longbibliography]{revtex4-1}
\usepackage{
  amsmath,
  booktabs,
  braket,
  graphicx,
  hyperref,
  longtable,
  nicefrac,
  siunitx,
  xcolor,
  xspace
}

\hypersetup{colorlinks,urlcolor=blue ,citecolor=blue ,linkcolor=blue}
\def\Er#1 {\textsuperscript{#1}Er}
\def\IP{\text{IP}}
\def\unit #1 #2 {\SI{#1}{#2}\xspace}
\def\wn#1{#1\,\text{cm}\textsuperscript{-1}}

\def\ts#1{\textsuperscript{#1}}
\def\tu#1{\textsubscript{#1}}

\date{Jul 2021}

\begin{document}
	\title{Spectroscopy of Rydberg States in Erbium using Electromagnetically Induced Transparency}
	\author{A.\,Trautmann}
	\affiliation{Institut f\"ur Quantenoptik und Quanteninformation, \"Osterreichische Akademie der Wissenschaften, Technikerstra{\ss}e 21a, 6020 Innsbruck, Austria}
	\author{M.\,J.\,Mark}
	\affiliation{Institut f\"ur Quantenoptik und Quanteninformation, \"Osterreichische Akademie der Wissenschaften, Technikerstra{\ss}e 21a, 6020 Innsbruck, Austria}	
	\affiliation{Institut f\"ur Experimentalphysik, Universit\"at Innsbruck, Technikerstra{\ss}e 25, 6020 Innsbruck, Austria}
	\author{P.\,Ilzh\"ofer}
	\altaffiliation[Present address: ]{5. Physikalisches Institut and Center for Integrated Quantum Science and Technology, Universit\"{a}t Stuttgart, Pfaffenwaldring 57, 70569 Stuttgart, Germany}
	\author{H.\,Edri}
	\author{A.\,El Arrach}
	\author{J.\,G.\,Maloberti}
	\affiliation{Institut f\"ur Quantenoptik und Quanteninformation, \"Osterreichische Akademie der Wissenschaften, Technikerstra{\ss}e 21a, 6020 Innsbruck, Austria}
	\author{C.\,H.\,Greene}
	\author{F.\,Robicheaux}
	\affiliation{Department of Physics and Astronomy, Purdue University, West Lafayette, IN 47907, USA}
	\affiliation{Purdue Quantum Science and Engineering Institute, Purdue University, West Lafayette, Indiana 47907, USA}
	\author{F.\,Ferlaino}
	\affiliation{Institut f\"ur Quantenoptik und Quanteninformation, \"Osterreichische Akademie der Wissenschaften, Technikerstra{\ss}e 21a, 6020 Innsbruck, Austria}	
	\affiliation{Institut f\"ur Experimentalphysik, Universit\"at Innsbruck, Technikerstra{\ss}e 25, 6020 Innsbruck, Austria}

\date{\today}
	
\begin{abstract}
 We present a study of the Rydberg spectrum in \ts{166}Er for series connected to the $4f^{12} (^3H_6) 6s$, $J_c=13/2 $ and $J_c=11/2 $ ionic core states using an all-optical detection based on electromagnetically induced transparency in an effusive atomic beam. Identifying approximately 550 individual states, we find good agreement with a multi-channel quantum defect theory (MQDT) which allows assignment of most states to $ns$ or $nd$ Rydberg series. We provide an improved accuracy for the lowest two ionization thresholds to $E_{\IP, J_c = 13/2 } = \wn{49260.750(1)}$ and $E_{\IP, J_c = 11/2 } = \wn{49701.184(1)}$ as well as the corresponding quantum defects for all observed series. We identify Rydberg states in five different isotopes, and states between the two lowest ionization thresholds. Our results open the way for future applications of Rydberg states for quantum simulation using erbium and exploiting its special open-shell structure.
\end{abstract}

\maketitle

\section{Introduction}
Rydberg states in neutral atoms have been highly successful to realize strongly interacting many-body quantum platforms. The large dipole-dipole (van der Waals) interaction, typically exceeding the MHz scale,  enables unique paths for quantum information processing and many-body quantum simulations \cite{Jaksch2000,Brennen2000,Saffman2010}.  
The conditions of strong interaction, long internal-state coherence, and microscopic control  can now be satisfied simultaneously when driving Rydberg (Ry) excitations in tweezer-trapped neutral atoms \cite{Browaeys2020}. This important advance was initially developed with alkaline atoms \cite{Barredo2016,Bernien2017,Barredo2018,Schmymik2020}. Due to their relatively sparse atomic structure, alkali are well suited for the implementation of robust, yet simple, cooling and trapping methods. However, this simplicity comes at the cost of somehow restricted opportunities for state  preparation and manipulation of the available internal degrees-of-freedom.

Recently, there has been a growing interest in extending the Rydberg toolbox to more complex multi-valence-electron atomic species, opening new possibilities for e.g. laser cooling and trapping, high-fidelity optical read-out, and quantum-information storage \cite{Saffman2008,Mukherjee2011,Topcu2014,RBS2018}. In multi-valence-electron atoms, the key paradigm shift is that, after Ry excitation of one of the available valence electrons, the core remains optically active, effectively resembling a single-charged positive ion.  
This active core leads for instance to a comparatively large polarizability \cite{Topcu2014}, allowing for optical trapping despite the repulsive ponderomotive potential for the Ry electron \cite{Wilson2019}. Remarkable progress has been made with two-electron atoms, i.e. the alkaline-earth Sr \cite{Millen2010,Bounds2018,Camargo2018,Couturier2019} and the alkaline-earth-like Yb atoms \cite{Lehec2018}, for ground-state tweezer trapping \cite{Norcia2018,Cooper2018,Saskin2019} and  more recently the combination of tweezer trapping and Ry excitation \cite{Wilson2019,Madjarov2020}.  

Pushing the boundaries even further, the next step is to consider atomic species with more than two valence electrons, like open-shell lanthanides for which laser cooling and quantum degeneracy have been demonstrated \cite{McClelland2006,Lev2010,Frisch2012,Miao2014,Cojocaru2017}.
Compared to alkaline-earth, these species might exhibit an ionic-core polarizability resembling even more the ground-state polarizability, and might allow access to a large hyperfine manifold \cite{Saffman2008,Robicheaux2018}. Beside the plethora of laser cooling transitions, they could also allow direct access to high orbital-momentum Ry states with negligible quantum defects, which are expected to have strong pair interactions because of the multitude of nearby degenerate quantum states. However, being a comparatively new quantum resource, open-shell Ry lanthanides remain rather unexplored in the ultracold regime.

So far, the only reported high-resolution Ry spectroscopy in ultracold open-shell lanthanides has been performed with Holmium in a magneto-optical trap \cite{Hostetter2015}. For Dy and Pm, resonance ionization spectroscopy (RIS) has been performed on a hot atomic vapor \cite{Haijun1992,Studer2016,Studer2019}. Using RIS, preliminary data for erbium are also available \cite{Studer2015}.

The present work reports on the first high-resolution Ry spectroscopy of erbium atoms. Using a two-photon scheme based on electromagnetically induced transparency (EIT) \cite{Boller1991,Mohapatra2007,Mauger2007,Naber2017}, we observe the $ns$ and $nd$ Rydberg series with principal quantum number $n$ ranging from 15 up to 140.  Using slope minimizing fits in Lu-Fano style plots \cite{LuFano1970}, we provide an improved value of the first ionization potential.
Our method does not require the magnetic quadrupole field present in a MOT, and thus allows an effortless high-resolution study of Zeeman shifts of the Rydberg states.  We use these shifts to assign the total angular momentum, $J$, of a subset of Ry $ns$ and $nd$ states, which serves as an important input parameter for the modelling of the Rydberg series.

Moreover, we identify characteristics of the Rydberg states using procedures based on Multichannel Quantum Defect Theory (MQDT), similar to recent work on Sr \cite{Vaillant2014} but accounting for Rydberg perturbers from a spin-orbit split threshold. Using the approximation that the $ns$ and $nd$ states do not mix at all, we successfully perform a MQDT fit to the $ns$ series. For the $nd$ series we introduce two different approximate methods which are fairly successful in representing most of the states but are less precise in representing the $nd$ perturbers or the measured g-factors.

Finally, we surprisingly observe that using just two-photon transition we could presumably couple ground-state atoms to an $ng$ Ry state ($\Delta \ell = 4$) thanks to the interaction between the submerged shells in Lanthanides. This result provides a first example of the uniqueness of Ry lanthanide, with respect to alkali and alkaline-earth atoms.

\begin{figure*}[th]
    \includegraphics{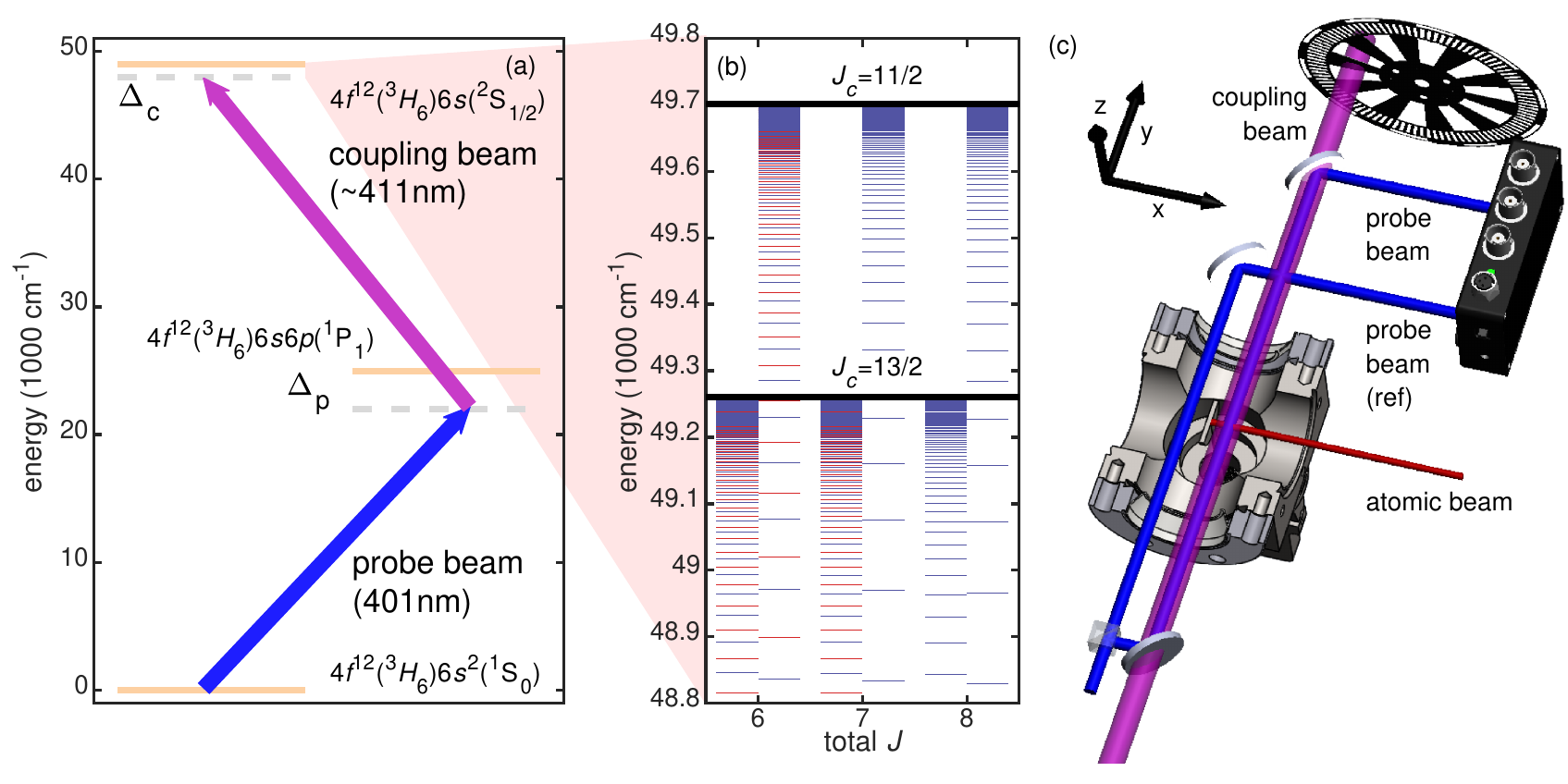}
    \caption{Electronic levels and experimental setup. (a) Excitation scheme involving the probe transition at about \unit 401 nm with single photon detuning $\Delta_p$ and the coupling transition around \unit 411 nm . Also shown the two-photon detuning $\Delta_c$. (b) Schematic Rydberg level scheme of the two lowest ionization thresholds for different total $J$. Red (blue) lines indicate the $ns$ ($nd$) series. (c) Schematic drawing of the experimental setup with the vacuum apparatus, atomic beam, and probe, reference, and couple laser beams.}
    \label{fig:sketch}
\end{figure*}

The paper is structured in the following way: Section \ref{sec:levelandsetup} discusses the energy levels and coupling schemes of erbium with all relevant states for this work. The experimental setup and measurement techniques are reviewed in Sec.~\ref{sec:setup}. In Sec.~\ref{sec:analysis} we will discuss the experimental data by demonstrating the EIT signal, and present the determination of $J$ and $g_J$ for a set of $nd$ states as well as an accurate value for the ionization threshold. Section \ref{sec:mqdt} presents our MQDT results for both $ns$ and $nd$ series and will compare the results with our experimental results.

\section{Considerations on the Erbium Level structure}
\label{sec:levelandsetup}

Figure~\ref{fig:sketch}(a)  shows the excitation scheme used in this manuscript and Fig.~\ref{fig:sketch}(b) a zoom-in onto the most relevant Rydberg series attached to the lowest two ionization thresholds. 
The electronic ground-state configuration of erbium reads as ($4f\ts{12}(\ts3H_6)6s\ts2(\ts1$S$_0$))\tu6. This type of configuration is often called sub-merged because the partially-filled inner f-shell is surrounded by the outer s-shell. Erbium has a total of 14 valence electrons, and each of those can be excited to a higher-lying state. The interactions inside the submerged shell determines the type of angular momentum coupling of the state. Generally speaking, the spin-orbit coupling scheme depends on the specific configuration and on the energy of the state --~ i.e.~overlap between the submerged wave-functions. 

For our two-photon Rydberg spectroscopy, we couple the ground-state to the intermediate state ($4f\ts{12}(\ts3H_6)6s6p(\ts1$P$^\circ\tu1))\tu7$. Here, the angular momentum coupling results from the small size of the $4f$-orbital compared to that of the $sp$ electrons. This leads to the $4f\ts{12}$ electrons coupling together to give a particular angular momentum term, ($\ts3H_6$), and the $sp$ electrons coupling to give ($\ts1$P$^\circ\tu1$). These two partial-$J$'s are then coupled to give the total $J=7$ for the intermediate state. A similar argument holds for the ground and first excited state of the ion ($4f\ts{12}(\ts3H_6)6s(\ts2$S$_{1/2}$))\tu{$J_c$} with $J_c=13/2$ for the ground ionic state and $J_c=11/2$ for the first excited state. For the Rydberg states, the energy scale of the Rydberg electron is the smallest suggesting the angular momentum of the positive ion should be coupled to that of the Rydberg electron. There are three ways to order the addition of angular momenta, but the interaction between the Rydberg states will typically preclude the states from being nearly pure in one ordering or the other. The order used in the MQDT simulations is to add the spin and orbital angular momenta of the Rydberg electron together to get the total angular momentum, $j$, of the Rydberg electron; this is then added to the total angular momentum of the core electrons to get the total angular momentum of the final state. Our two photon excitation scheme leads to even parity Rydberg states with most of the states having $ns$ or $nd$ character. Also, Lanthanides offer the unique possibility to directly couple to $ng$ states with $\ell = 4$ due to the angular momentum of the submerged shell. Table \ref{tab:states} lists the relevant quantum numbers for the states discussed in this paper.

\begin{table}[ht]
\begin{ruledtabular}
\caption{Quantum numbers of all relevant states for the EIT scheme, with the total ground state, intermediate state, and ion core states with Rydberg series attached to. The table lists the configuration, state energy, as well as total angular momentum $J$. For Rydberg states, also the core total angular momentum $J_c$ is given. Only states with \mbox{$|\Delta J| \leq 1$} with respect to the intermediate state ($J_{ex}=7$) are shown as our two-photon scheme only couples to such states.}
\label{tab:states}
\begin{tabular}{lll}
configuration & term, $J$ & \kern-0cm \parbox{1.9cm}{energy (\!\wn{}\!)\\ (threshold)} \\
\colrule
    $4f^{12}(^3H_6)6s^2{}$($^1$S$_0$) &  (6,0) 6 & 0\\
    $4f^{12}(^3H_6)6s6p{}$($^1$P$^\circ_1$) &  (6,1)$^\circ$ 7& 24943.298 \\
    $4f^{12}(^3H_6)6s _{1/2}$ &  (6,1/2) 13/2 & 49260.750\\
    $4f^{12}(^3H_6)6s _{1/2}$ &  (6,1/2) 11/2 & 49701.184\\
\colrule
    $4f^{12}(^3H_6)6s _{1/2}ns_{1/2}$ & (13/2,1/2)6,7 & ($J_c = 13/2$) \\
    $4f^{12}(^3H_6)6s _{1/2}ns_{1/2}$ & (11/2,1/2)6 & ($J_c = 11/2$) \\
\colrule
    $4f^{12}(^3H_6)6s _{1/2}nd_{3/2}$ & (13/2,3/2)6,7,8 & ($J_c = 13/2$)\\
    $4f^{12}(^3H_6)6s _{1/2}nd_{5/2}$ & (13/2,5/2)6,7,8 & ($J_c = 13/2$)\\
    $4f^{12}(^3H_6)6s _{1/2}nd_{3/2}$ & (11/2,3/2)6,7 & ($J_c = 11/2$)\\
    $4f^{12}(^3H_6)6s _{1/2}nd_{5/2}$ & (11/2,5/2)6,7,8 & ($J_c = 11/2$)\\
\end{tabular}
\end{ruledtabular}
\end{table}

\section{Experimental setup}
\label{sec:setup}

Figure~\ref{fig:sketch}(c) shows a schematic drawing of the experimental spectroscopy setup, consisting of an ultra-high vacuum setup with a high-temperature effusion cell, a transversal cooling chamber, a differential pumping section (not shown here), and the probe chamber. The design is similar to the one in~\cite{Ilzhofer2018}. Erbium atoms are evaporated in the effusion cell at $1300^\circ$C. From the effusion cell, the atoms pass through three apertures to form a collimated beam, propagating along the horizontal $x$-direction. In the subsequent probe chamber, the atomic beam crosses the interaction region with the coupling and probe laser beam. The coupling and probe beam counter-propagate and intersect the atomic beam perpendicular~($y$-direction) to reduce Doppler shifts. They are overlapped and separated by dichroic mirrors. An additional reference beam, split from the probe beam, propagates parallel to the probe beam and acts as a reference to cancel out power fluctuations.  We can block the atomic beam between the probe and reference beam to provide a reference with and without the atomic absorption. We use about 10 to \unit 500 {\micro W} , with a waist ($1/e^2$ radius) of about \unit 0.5 mm for both the probe and reference beam, while the coupling beam approximately has \unit 130 mW at the interaction region, and a waist of  \unit 1 mm . We modulate the coupling beam with an optical chopper at \unit 7 kHz , monitor both the probe and reference beam on balanced photodiodes (PD1 and PD2), and feed the AC-coupled difference of these signals as input for a lock-in amplifier. Both lasers are commercial resonantly frequency-doubled devices; the probe and reference beam are derived from an amplified diode laser locked onto the \unit 401 nm transition line using a Doppler-free modulation transfer spectroscopy in a hollow cathode lamp~\cite{Frisch2012}. The coupling laser is derived from a continuous-wave free-running Ti:Sa laser. Both lasers are monitored on a wavelength meter with \unit 60 MHz absolute accuracy, see App.~\ref{Appendix:Error}.

We add quarter- and half-wave plates to the probe and coupling beam path to control the light polarization, and apply a magnetic field on the order of \unit 10 G in~$z$ ($B = (0, 0, B_z) $) direction. This enables us to drive $\sigma^{\pm}$ and/or $\pi$ transitions in a controlled setting, which facilitates the assignment of the total angular momentum $J$ and estimation of the g-factor of Rydberg states, see Sec.~\ref{sec:ExpJ}.

\section{Data Analysis}
\label{sec:analysis}
\begin{figure}[ht]
    {
    \includegraphics{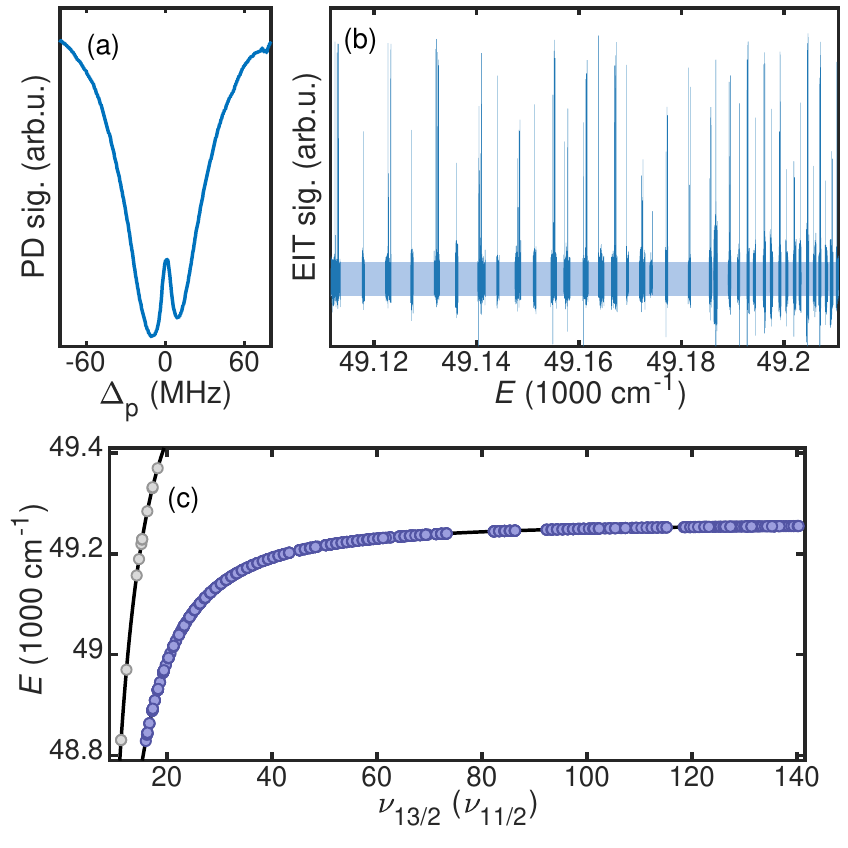}}
    \caption{EIT spectroscopy and survey of Rydberg states. (a) Exemplary EIT resonance around \wn{49147.967}. Here, the coupling beam is kept fixed while $\Delta_p$ is scanned, showing the typical EIT signal of reduced absorption when on 2-photon resonances. (b) EIT spectroscopy over a broad range of energies covering about $100$ individual Rydberg states. $\Delta_p$ is fixed to zero while the coupling laser frequency is scanned. The shaded area indicates the noise floor as a guide to the eye. (c) Energy of all observed Rydberg states extracted from the data as a function of the assigned effective principle quantum number. Ry states that are either above the first ionization threshold or comparatively broad are assigned to the second-lowest $J_c = 11/2$ ionization threshold and are plotted against their corresponding principle quantum number (grey). Solid lines show the expected Rydberg energies using the simple Rydberg formula with the derived $E_{\text{ion},j}$.}
    \label{fig:traces}
\end{figure}

To detect the Rydberg levels, we make use of electromagnetically induced transparency \cite{Boller1991}. In short, we detect the probe laser transmission through the atomic beam, which experiences absorption when in resonance with the atomic transition to the intermediate state. In case the coupling laser hits a resonance condition i.e. couples the intermediate state to a Rydberg state, this absorption gets reduced. This can be understood in a dressed state picture where the coupling leads to a doublet of dressed states (Autler–Townes doublet) \cite{Autler1955}, and together with destructive interference of the absorption of these states, to a transparency window at resonance. This tell-tale sign of EIT is shown in Fig.~\ref{fig:traces}(a) where we directly record the probe laser power after passing the atomic beam. The probe laser frequency is scanned over the absorption resonance, while the coupling laser frequency is fixed, in this case on resonance to the Rydberg state at \wn{49147.967}, identified as the lowest-lying fine-structure state of the $31d$ Rydberg manifold, see later discussion. The narrow transmission peak due to EIT appears in the center of the absorption line.

For a survey of Rydberg states, we lock the probe laser onto the hollow-cathode lamp spectroscopy on resonance with the \unit 401 nm transition to the intermediate state, scan the coupling laser frequency and again record the transmitted probe laser power. We additionally improve our Signal-to-Noise ratio by using a reference beam, which we subtract to reduce noise from power fluctuations, and using a lock-in technique where we modulate the coupling beam with a chopper, see Sec. \ref{sec:levelandsetup} for details. In case we hit the resonance condition to a Rydberg state, we observe directly the increased transmission. Figure~\ref{fig:traces}(b) shows an excerpt of the total spectroscopy data in the region from about \wn{49112} to \wn{49211}.

Using this technique, we record about 550 EIT resonances, and assign their total energy, see App.~\ref{sec:allstates} for a full list. After the determination of the ionization thresholds (see details in Sec.~\ref{sec:threshold}) we can assign effective quantum numbers to each level. Figure~\ref{fig:traces}(c) shows the energy of all observed Rydberg states as a function of their effective quantum numbers. We observe the typical $1/n^2$ scaling of Rydberg states. We also found several very strong EIT features, at least ten times stronger than any surrounding resonances. Together with a few states located above the first ionization threshold, we assign them to be part of the Rydberg series attached to the second-lowest $J_c = 11/2$ ionization threshold. Also here we observe a similar $1/n^2$ scaling which, together with their positions agreeing with the expected locations of states from this threshold, further strengthens our assignment.

\subsection{Determination of lowest ionization threshold and assignment of series}
\label{sec:threshold}
As an important parameter for the assignment of quantum numbers and the understanding of the Rydberg series, we first determine the lowest ionization threshold, by plotting the effective quantum numbers \mbox{$\nu_j = \sqrt{R_\mu/(E_{\text{ion},j} - E_i)}$} and effective quantum defects $\mu_j = -\nu_j$(mod 1). Here, $\nu_{j}$ is the effective quantum number with respect to the ion core state with angular momentum $j$, $E_{\text{ion},j}$ the corresponding ionization threshold, and $E_i$ the energy of the Rydberg states. We also use $n_\text{eff} = \rm{floor}(\nu_j)$ as the integer part for the assignment. We use a Lu-Fano analysis of our data to extract a new value for the ionization threshold: For an unperturbed Rydberg series, the quantum defect is nearly constant for intermediate  principle quantum numbers. For Rydberg states with very high principle quantum numbers close to the ionization threshold, external influences like electric fields can disturb the states, and uncertainties in absolute frequencies have a larger influence on the effective quantum defect, while at lower energies the quantum defect shows a stronger state dependence. By plotting the calculated effective quantum defect $\mu_{13/2}$ versus $\nu_{11/2}$, we obtain a manifold of flat series around $\mu_{13/2} = 0.8$.
We vary the value for the ionization threshold and fit a straight line to all states with $0.7 < \mu_{13/2} < 0.9$ above \wn{49250}, and find $E_{\text{ion},13/2} = \wn{49260.7442(23)}$ as a value for the lowest ionization threshold to minimize the overall slope. This value is within the error margin of the literature value of \wn{49262(8)} \cite{Worden1978} and of the preliminary value \wn{49260.73(9)} derived in Ref.~\cite{Studer2015}, but improves in precision by almost four/two orders of magnitude. For the first excited ionization threshold, the same analysis suffers from the low number of states and missing states at high $\nu_{11/2}$. Therefore, we use the value reported in \cite{Martin1978,Wyart2009} for the splitting of the two states with \wn{440.433(10)}, and calculate the $E_{\text{ion},11/2} = \wn{49701.177(10)}$.

\begin{figure}[t]
    \centering
    \includegraphics{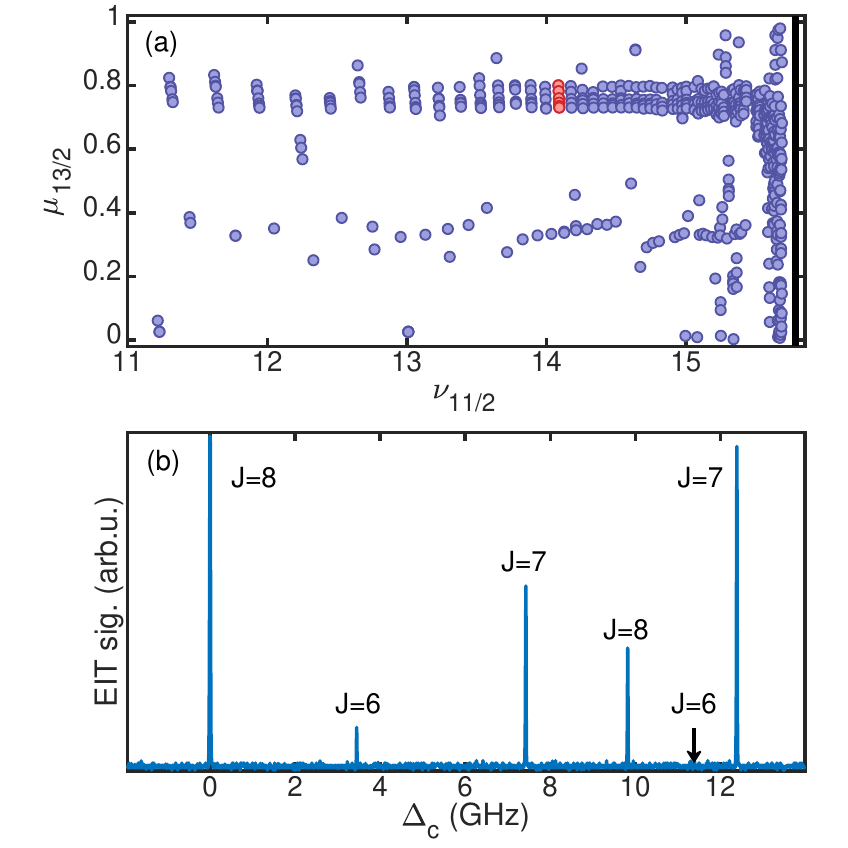}
    \caption{Rydberg series: (a) Lu-Fano-style plot, showing the calculated $\mu_{13/2}$ of all observed states below the lowest ionization threshold as a function of the effective quantum number $\nu_{11/2}$ of the second-lowest ionization threshold. (b) Zoom into the EIT spectrum around \wn{49148}, showing the states marked as red datapoints in (a). This bundle of resonances is identified as the $31d$ Ry state where the fine structure splitting leads to six features with \mbox{$J = 6, 7$, or $8$}, see Sec. \ref{sec:ExpJ} for the assignment.}
    \label{fig:alldata}
\end{figure}

Figure~\ref{fig:alldata}(a) shows the resulting effective quantum number and quantum defect for all states below the first ionization threshold. Two main series are visible, the first one around $\mu_{13/2}=0.35$ which consists of single separated states, expected for the $ns$-series. At the second series around $\mu_{13/2}=0.75$ we observe a bundle of $5-6$ lines in relatively close proximity to each other. This is consistent with the expected $nd$-series. Figure~\ref{fig:alldata}(b) shows an exemplary scan over one of those bundles and gives already the assignment of the individual $J$-values, as we will detail below. 

\subsection{Determination of total angular momentum \textit{J} and g-factors}\label{sec:ExpJ}
While the $ns$ states are clearly separated and are expected to be nearly exclusively $J=6$ (see Sec. \ref{sec:mqdt} for a more detailed discussion), there is a large number of possible $nd$ states. A full assignment of the $nd$ states requires a determination of their total angular momentum, $J$, which we extract by performing a Zeeman spectroscopy for each state of the fine structure manifold for principal quantum numbers $n_\text{eff} = 23, 27, 31, 37$: We apply a magnetic field along $z$, and use horizontal polarization for all laser beams (propagating along $y$, polarization parallel to $x$), which provides $\sigma^+$ and $\sigma^-$ light in the reference frame of the atoms. We first calibrate our magnetic field with a Zeeman spectroscopy of the transition to the intermediate state without probe light by scanning the \unit 401 nm laser and fitting gaussian curves to the absorption signal. With the known g-factors for both ground and intermediate state we determine our magnetic field strength to be \unit 10.7(5) G .

\begin{figure}[t]
    \centering
    \includegraphics{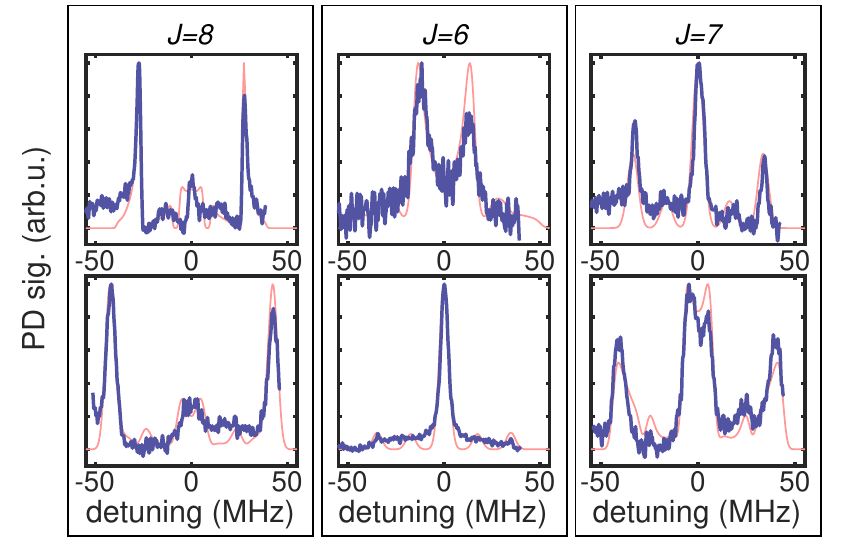}
    \caption{Zeeman mapping of the $27d$ fine structure multiplet. Blue thick lines show the photo diode signal of multiple averaged experimental data traces scanning the couple laser frequency. Red thin lines show the estimated spectral pattern using the fitted g-factors.}
    \label{fig:zeeman}
\end{figure}

Now we lock the probe beam frequency again on the zero-field resonance, scan the couple laser frequency over a range of about \unit 100 MHz , and observe a splitting of the EIT lines, which results in distinctly different signal patterns, see exemplary the patterns of the $27d$ fine structure manifold shown in Fig. \ref{fig:zeeman}. We can sort the patterns into three groups: (A) One weak central peak and two strong peaks shifted symmetrically by about \unit 40 MHz , (B) one central peak which may split into two peaks separated by less than \unit 30 MHz , (C) a strong central peak with two or four weaker side peaks.

These three distinct behaviors can be explained as the result of the combination of the specific $J$ in the ground ($J_{gs}$), excited ($J_{ex}$) and Rydberg states and the polarization of the probe and coupling beam. In general, the resonance frequency of a specific 2-photon transition $|J_{gs},m_{gs}\rangle\rightarrow|J,m_{J}\rangle$ shifts from its zero field energy as $\Delta E = (g_{J}m_{J}-g_{gs}m_{gs})\mu_B B$. Here, $m_{gs}$ ($m_{J}$) denotes the projection of $J_{gs}$ ($J$) of the groundstate (Rydberg) atom along the quantization axis and $\Delta m=m_{J}-m_{gs}$ the total difference between them. As transitions are limited to $|\Delta m|\leq 2$ due to selection rules and assuming that $g_{J}$ is close to $g_{gs}$ we can identify $5$ main features with energies:
\begin{align}
\Delta E_{\pm2} &= \pm2g_{gs}\mu_B B & (\Delta m&=\pm2) \nonumber \\
\Delta E_{\pm1} &= \pm1g_{gs}\mu_B B & (\Delta m&=\pm1) \nonumber \\
\Delta E_{0} &= 0 & (\Delta m_J&=0) \nonumber
\end{align}
Taking into account the additional difference in g-factors, each of these main resonances splits again into a series of closely spaced resonances with additional energy shifts $\Delta E = \Delta g_Jm_{J}\mu_B B$, with $\Delta g_J=g_{J}-g_{gs}$.

Given our applied polarization of probe and coupling beam ($\sigma^\pm$) and partial optical pumping towards the stretched states $J_{gs}=\pm6$ during the spectroscopy due to the different Clebsch-Gordan coefficients of the first transition we expect mainly resonances at $\Delta E_{\pm2}$ and $\Delta E_{0}$, see App.~\ref{Appendix:Specsim} for further details and corresponding calculations. Based on these modelled spectral patterns, for $J=8$ we expect the strongest signals at $\Delta E_{\pm2}$  and a weaker central peak at $\Delta E_{0}$. Instead, for $J=6$, the strongest resonance will be $\Delta E_{0}$ which might split for large $\Delta g_J$, while $\Delta E_{\pm2}$ transitions will be very weak. Finally, $J=7$ will have its strongest component at $\Delta E_{0}$ together with slightly weaker $\Delta E_{\pm2}$ transitions and some very weak $\Delta E_{\pm1}$ components. With these considerations we assign group (A) to $J=8$, group (B) to $J=6$ and group (C) to $J=7$.

Using this assignment technique for all investigated principal quantum numbers $n_\text{eff} = 23, 27, 31, 37$~\footnote{For $n_\text{eff} = 23$ we observe two $J=8$ features very close together, which is not yet understood.}, we observe the same fine-structure pattern with increasing energy, which goes as $J=8,6,7,8,6,7$, see Fig.~\ref{fig:alldata}(b). Additionally, we find that the $J=6$ EIT resonances within one scan are significantly weaker compared to $J=7,8$. The same pattern in relative signal strength can be found for five other $n_\text{eff} = 32-36$, allowing us to assign the total $J$ of each state via comparison of the order and relative height of the resonances of the EIT spectra without the need of a full Zeeman mapping for every $n_\text{eff}$.

We are also able to experimentally determine the $g_J$ value for most of the investigated states. For this, we fit our modelled spectral pattern (see App.~\ref{Appendix:Specsim}) to the experimental data, with $g_J$ as the main fitting parameter. We also allow some variation of the polarization and the optical pumping effect to be able to account for differences in the experimental conditions like probe and coupling laser intensities and beam alignment. Our results are summarized in Table~\ref{tab:gfactor}. 

\section{MQDT, assignment of lines, assignment of quantum defects}
\label{sec:mqdt}

Most of the experimentally measured energies for $20<\nu_{13/2}<60$ seem to be grouped into two sets that are weakly interacting. The group with quantum defects $0.3\lesssim\mu_{13/2}\lesssim0.4$ appear to come from a two channel series with one channel attached to the $13/2$ threshold and the other attached to the $11/2$ threshold. From the discussion in App.~\ref{sec:mqdtApp}, this suggests that these are $J=6$ states with $ns$ Rydberg character. The other main group are states with quantum defects between $\sim 0.75$ and $\sim 0.8$ which, due to their number, must be states with $nd$ Rydberg character. There are several states with small quantum defects which might have $ng$ character. Finally, there are several states that are not part of either group but might be identified by fitting the MQDT parameters.

\subsection{\textit{ns}-series}\label{Secns}

In the limit that coupling between $ns$ and $nd$ states can be
ignored, the $ns$-series with $J=6$ results from a two channel
system with one channel attached to the $J_c=13/2$ threshold and
one attached to $J_c=11/2$. Since the $K$-matrix is symmetric
there are only 3 independent parameters in the $K$-matrix. We chose
the parameters to be the two eigen-quantum defects and the mixing
angle of the eigen-vector. Taking channel 1 to be $J_c=13/2$
and 2 to be $J_c=11/2$, the $K$-matrix is written as
\begin{equation}
K_{ij}= \sum_a U_{ia}U_{ja }\tan (\pi\mu_a )
\end{equation}
where $U_{11}=U_{22}=\cos (\alpha )$ and $U_{21}=-U_{12}=\sin (\alpha )$.
The frame transformation approximation in App.~\ref{SecFT} implies
$\alpha = \cos^{-1}(\sqrt{7/13})=0.7469$.

\begin{figure}[ht]
    \centering
    \includegraphics{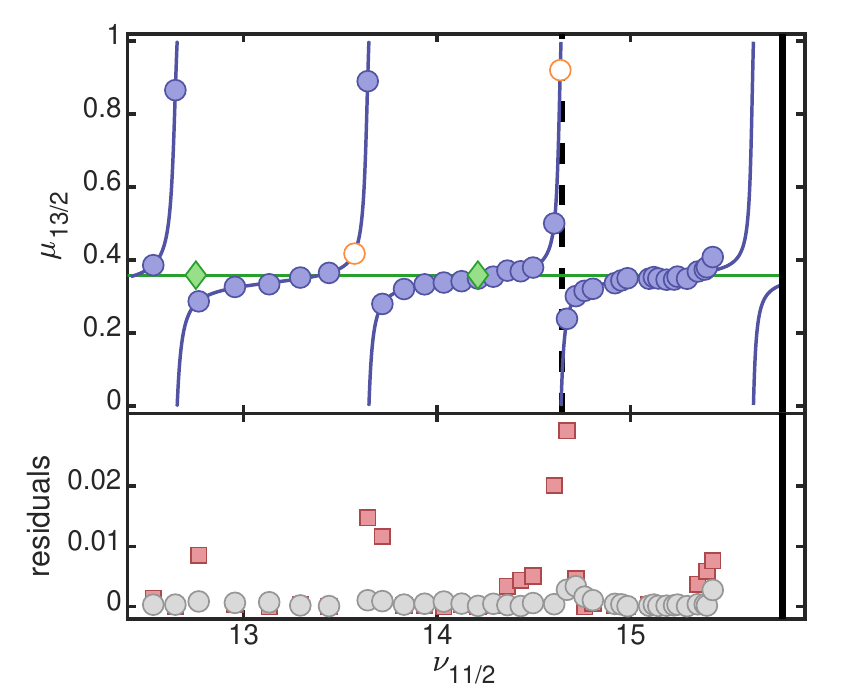}
    \caption{Lu-Fano plot of the $s$ series states. (a) All experimentally observed states assigned to the $J = 6$ (blue circles) and $J = 7$ (green diamonds) $s$ series attached to the lowest $13/2$ ionization threshold. The lines show the result of the MQDT calculations with energy dependence for the $ns$ series; the black bar indicates the ionization threshold. The dashed line indicates a broad and strong EIT feature, assigned to the $ns$ series of the $11/2$ threshold. The open orange circles indicate theoretically predicted $J = 6$ states found experimentally in an additional survey. (b) Absolute values of the difference between the measured data and the MQDT model, with (grey circles) and without (red squares) energy dependence (see Sec.~\ref{Secns} for details on the model).}
    \label{fig:lufano}
\end{figure}

We fit the identified states using these 3 parameters. The fit minimized the $\chi^2$ calculated from the sum of differences between the calculated and measured energies. When we take these parameters to be independent of energy, we find $\mu_1=0.27458$, $\mu_2=0.42223$, and $\alpha =0.72885$. Note the closeness of $\alpha$ to the value expected from the frame transformation approximation which further supports this approximation for the $ns$-series. The residuals of the $\mu_{13/2}$ are shown in Fig.~\ref{fig:lufano}(b) and for most of the states they are smaller than $\sim 0.01$. This level of agreement is roughly what should be expected for states over such a large range of energy. For a more realistic fit, we included a linear energy dependence in all of the parameters. Historical precedence gives a subscript ``2" to the linear energy dependence (because the linear energy dependence was originally written as $1/(n-\mu )^2$). For a parameter $\tau$, the energy dependence is written as
\begin{equation}
\tau (E)=\tau_0 +\eta (E)\; \tau_2
\end{equation}
where $\eta (E)= (E_{13/2}-E)/R_{166}$ and $R_{166}$ is the Rydberg constant defined in App~\ref{sec:mqdtApp}. The fit values are given in Table~\ref{Tabns}. Note that the value for $\alpha$ is only $\sim 3$\% different from the frame transformation value. The comparison between the MQDT fit and measured bound state energies is shown in Fig.~\ref{fig:lufano}. Note that the residuals for all states are now much less than $0.01$. The fit was useful in identifying states not obviously part of the series with quantum defects between $\sim 0.3$ to $\sim 0.4$; these are the 3 states with quantum defects $\sim 0.9$. The fit also predicted two states not in the original data set that were subsequently identified: $26s$ at \wn{49105.45(7)} ($\mu_{13/2}\simeq 0.42$) and $39s$ at \wn{49188.91(7)} ($\mu_{13/2}\sim 0.92 $).

\begin{table}[ht]
\begin{ruledtabular}
    \centering
    \caption{The MQDT parameters for the $ns$ $J=6$ series.}\label{Tabns}
    \begin{tabular}{lcc}
    $\tau$ & $\tau_0$ & $\tau_2$ \\
    \hline\hline
    $\mu_1$ & 0.30136 & -15.78 \\
    $\mu_2$ & 0.42673 & -1.718 \\
    $\alpha$ & 0.72378 & -0.594 \\
    \end{tabular}
\end{ruledtabular}
\end{table}

We observed only two states with quantum defects of $\mu = 0.358$ that do not appear to be members of the ${J=6}$ series. We have assigned these to the $J=7$ series. This small number of observed states compared to $J=6$ might be explained by the different coupling of the two series. At short range, the $ns$ $J=7$ must have the coupling $(6sns)\; ^3$S$_1$. Since the intermediate state has $(6snp)\; ^1$P$_1$ character, transitions to the $J=7$ would be dipole suppressed. The large difference of this quantum defect from the $\mu_2$ for $J=6$ suggests limitations to the frame transformation because these values should both be the $^3$S$_1$ quantum defect, App.~\ref{SecFT}.

\subsection{\textit{nd}-series}

The $nd$-series are more difficult to model because of the larger number of channels and the limited number of experimentally identified states being perturbed, see Sec.~\ref{sec:ExpJ} above. In total the experimental data allowed to identify 12 states with $J=6$, 18
states with $J=7$, and 19 states with $J=8$ character. These
states were used in the fit of the parameters. Because the $J=6$
and 7 have 4 channels and the $J=8$ has three channels, there are
10 free parameters for $J=6$ and 7 and 6 free parameters for $J=8$.
Symmetries in the bound state conditions mean that the energies only determine 8 parameters for $J=6$ and 7 and 5 parameters for $J=8$
(whenever 2 channels are attached to a threshold, the energies only
determine the eigen-quantum defects and can not determine their
mixing angle). The main difficulty in directly determining the full $K$-matrix
is that the $J_c=13/2$ $nd$ states experimentally identified to a particular $J$ are not strongly perturbed by states attached to the $J_c=11/2 $ threshold. This means that the eigen-quantum defects attached to the $J_c=13/2$ threshold are well defined but most of the $K$-matrix elements are relatively unconstrained.
As a contrast, see Fig.~\ref{fig:lufano} where the perturbations of the $ns$ states
clearly lead to variations greater than 0.1 for $\mu_{13/2}$.
This means the energies are giving information about 2 parameters for each $J$.

These considerations suggest using the frame transformation approximation or interaction approximation, discussed in Apps.~\ref{SecFT} and \ref{SecQP} respectively, to reduce the number of fit parameters. These approximations lead to nearly the same $\chi^2$ fit to the data points even though they have very different
physical motivation. This is because they are mainly fitting to the eigenquantum defects of the channels attached to the $J_c=13/2$ threshold. For these fits, we did not include energy dependence in the MQDT parameters because the resulting fits did not substantially improve the agreement with experimental results.

In addition to the measured energies, we were able to experimentally determine the g-factor of several of the $nd$ states. Section~\ref{Secgfac} describes an approximate method for calculating g-factors given the MQDT parameters. If the MQDT parameters were exact, this approximation would lead to errors less than 1\%.
Unfortunately, we did not find an effective way of using
the measured g-factors in the fitting procedure. The difference between the
measured and simulated g-factors described below results from the
limitations of the frame transformation and interaction approximations
used in the $nd$-fits.

\begin{figure}[ht]
    \centering
    \includegraphics[width=\columnwidth]{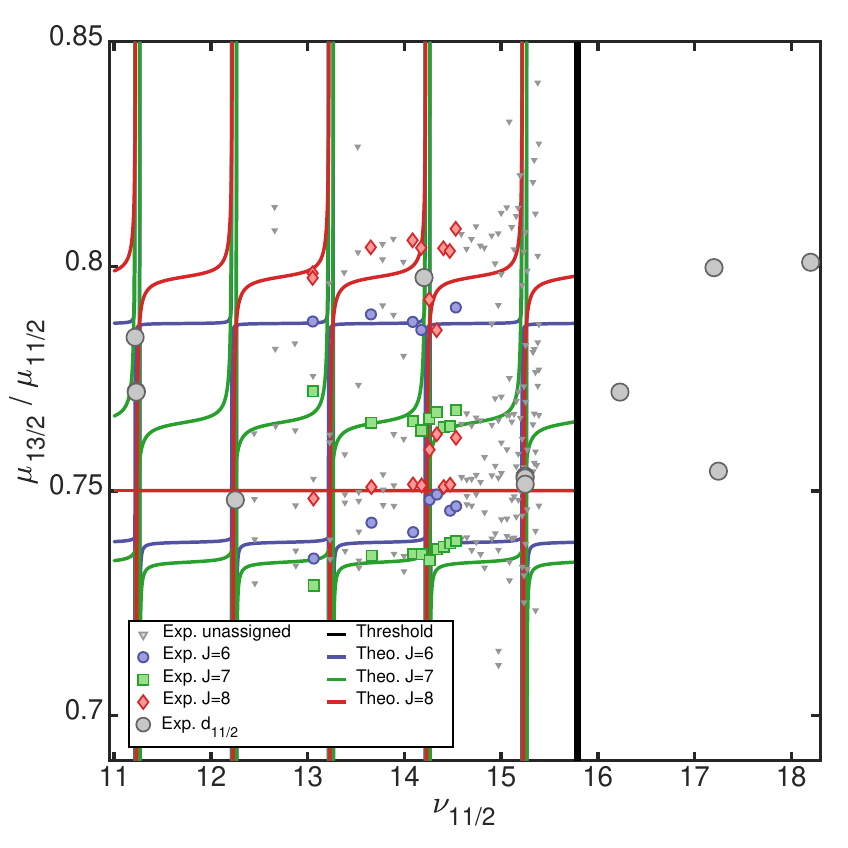}
    \caption{MQDT assignment of $nd$ Rydberg series. The lines show the results of the frame transformation approximation fitting. Small symbols show all experimental data, colored filled symbols show all states where we assigned $J$ experimentally. Large grey circles (plotted with their respective $\mu_{11/2}$) represent states which we assign to the $J_c = 11/2$ threshold. Four of these are above the first threshold, while states for four different $n$ are identified as perturbers of the $d$ series attached to the $J_c = 13/2$ threshold.}
    \label{fig:dseries}
\end{figure}

\begin{table}[ht]
\begin{ruledtabular}
    \centering
    \caption{The MQDT parameters for the $nd$ series for
    different $J$.}\label{Tabnd}
    \begin{tabular}{lcccc}
    $J$ & $\mu (^1D_2)$ & $\mu (^3D_1)$ & $\mu (^3D_2)$ & $\mu (^3D_3)$ \\
    \hline\hline
    6& 0.782 & 0.729 & 0.744 & 0.791 \\
    7& 0.752 & 0.724 & 0.747 & 0.794 \\
    8& 0.750 & \null  & 0.763 & 0.808 \\
    \end{tabular}
\end{ruledtabular}
\end{table}

Figure~\ref{fig:dseries} shows the experimental states where $J$ has been identified, together with the calculated $nd$-series using the frame transformation approximation . This plot emphasizes that the fit gives a fairly accurate representation of the experimentally detected states.
Table~\ref{Tabnd} shows the fit parameters for the
frame transformation approximation.
The fact that the quantum defects
mostly do not vary strongly with $J$ (only the $^1D_2$ varies by
more than 0.02) suggests that this approximation captures much of
the physics of these series. However, this approximation is not accurate
enough to  predict the states that are perturbed or
are attached to the $J_c=11/2$ threshold, see Fig.~\ref{fig:dseries}.
Also, the calculated g-factors substantially differ from the measured
values, see Tab.~\ref{tab:gfactor}.
Both difficulties suggest that the frame transformation does
not capture the full physics. The channel character is approximately
40:60 (or 60:40) of nd$_{3/2}$ and nd$_{5/2}$ for the
frame transformation approximation. Thus, this approximation suggests
the channels are strongly mixed.
The fit to the interaction parameters,
App.~\ref{SecQP}, gives $C_q=0.7533$, $\mu_{3/2}=-0.0777$, $\mu_6 = 0.7840$,
$\mu_7=0.7690$, and $\mu_8=0.7979$. Since the experimental states
are constraining 6 $K$-matrix parameters and there are 5 parameters
in the fit, the fact that this model accurately reproduces the
experimental values argues for the physics contained in the method.
However, this approximation is also not accurate enough to
predict the states that are perturbed or
are attached to the $J_c=11/2$ threshold, see Fig.~\ref{fig:dseries}.
It does a better job of predicting the g-factor but still has
substantial inaccuracies, see Tab.~\ref{tab:gfactor}. The channel
character is nearly pure for this approximation. However, the state
designation is reversed from that used to experimentally determine
the g-factor in Tab.~\ref{tab:gfactor}. This contradicts the frame
transformation characterization of the channel mixing which points
to the uncertainty in the modeling of the $nd$ series.

\begin{table}[ht]
\begin{ruledtabular}
    \centering
    \caption{List of fine structure states with Zeeman mapping and measured g-factor, determined by energy splitting in a vertical magnetic field. For some states the Signal-to-Noise ratio was too low to reliably fit the g-factor (--). The calculated g-factors are from the (Frame
    Transformation approximation, Interaction Fit). Note that the calculations only show a single state around 49056.9.}
    \def\ignoreiv#1#2#3#4{}
    \begin{tabular}{rc<{(5)}crrr}
    $n_\text{eff}$ & E (cm$^{-1}$)\ignoreiv & $J$ & $g_{J\text{, meas}}$ & $g_{J\text{, diag}}$ & $g_{J\text{, calc}}(FT,QP)$\\
    \hline\hline
    23 & 49056.886 & 8 & -- & -- & (--,--) \\
    23 & 49056.904 & 8 & 1.21(2) & 1.1493 & (1.2205,1.2256) \\
    23 & 49057.075 & 6 & 1.34(3) & 1.2454 & (1.1597,1.2259) \\
    23 & 49057.347 & 7 & 1.17(4) & 1.1877 & (1.1491,1.2262) \\
    23 & 49057.764 & 8 & -- & 1.2234 & (1.1520,1.1471) \\
    23 & 49057.997 & 6 & 1.23(5) & 1.2293 & (1.3148,1.2488) \\
    23 & 49058.102 & 7 & 1.33(9) & 1.2258 & (1.2640,1.1873) \\
    27 & 49112.369 & 8 & 1.10(2) & 1.1493 & (1.2206,1.2256) \\
    27 & 49112.531 & 6 & 1.32(2) & 1.2454 & (1.1597,1.2259) \\
    27 & 49112.794 & 7 & 1.13(6) & 1.1877 & (1.1492,1.2262) \\
    27 & 49112.949 & 8 & 1.23(7) & 1.2234 & (1.1520,1.1471)\\
    27 & 49113.035 & 6 & 1.17(8) & 1.2293 & (1.3149,1.2488) \\
    27 & 49113.115 & 7 & 1.24(10) & 1.2258 & (1.2642,1.1873) \\
    31 & 49147.967 & 8 & 1.18(5) & 1.1493 & (1.2206,1.2256) \\
    31 & 49148.098 & 6 & 1.33(2) & 1.2454 & (1.1597,1.2259) \\
    31 & 49148.258 & 7 & 1.22(4) & 1.1877 & (1.1491,1.2262) \\
    31 & 49148.359 & 8 & 1.30(3) & 1.2234 & (1.1520,1.1471) \\
    31 & 49148.435 & 6 & 1.17(5) & 1.2293 & (1.3149,1.2488) \\
    31 & 49148.470 & 7 & 1.25(3) & 1.2258 & (1.2641,1.1873) \\
    37 & 49181.406 & 8 & 1.13(8) & 1.1493 & (1.2207,1.2256) \\
    37 & 49181.480 & 6 & --     & 1.2454 & (1.1597,1.2259) \\
    37 & 49181.578 & 7 & 1.12(10) & 1.1877 & (1.1492,1.2262) \\
    37 & 49181.604 & 8 & 1.23(2) & 1.2234 & (1.1520,1.1471) \\
    37 & 49181.668 & 6 & --     & 1.2293 & (1.3149,1.2488) \\
    37 & 49181.701 & 7 & 1.22(3) & 1.2258 & (1.2642,1.1873) \\
    \end{tabular}
    \label{tab:gfactor}
\end{ruledtabular}
\end{table}

Note that while the two-threshold MQDT analysis appears to predict much of the general pattern of the Rydberg levels and perturbations at least qualitatively correctly, the energy range just below the $J_c=13/2$ threshold looks irregular and even rather chaotic.  To account for this apparent irregularity, it should be remembered that there are higher energy levels of Er$^+$ that are guaranteed to support their own infinite Rydberg series of levels as well.  It is expected that some of the lower-lying members of those Rydberg series can also occur in the spectral region studied here, and those are expected to cause significant distortions of the experimental energy level pattern and associated deviations from the present MQDT models.

One ionic threshold in particular appears to lie at an energy poised to produce such a perturbation just below the even-parity ground state of the ion.  We refer to the odd-parity level of Er$^+$ with the spectroscopic label $4f^{11}6s^2\ ^4$I$^o_{15/2}$, which lies \wn{6824.774} above the  $J_c=13/2$ ionic ground state.  A $5f$ Rydberg electron attached to that core, with an expected quantum defect of $\mu_f \approx 1$, should produce many Rydberg states of erbium lying approximately \wn{34} below the $J_c=13/2$ ionization threshold, in the $\nu_{13/2} \approx 50-60$ range, and $\nu_{11/2} \approx 15.2$.  In specific, a $5f_{5/2}$ electron produces levels with all values of $J$ from 5 to 10, and a $5f_{7/2}$ electron produces $J$ from 4 to 11, so there are 6 perturbing levels expected near that region of the spectrum with angular momenta in the range $J=6-8$ that would be observable in the present experiment.  Some of those levels might be present in Fig.~\ref{fig:dseries}, as that region near $\nu_{11/2} \approx 15.2$ shows numerous levels that deviate from our simplified two-threshold MQDT models.

\subsection{Coupling to low-lying \textit{ng} states}
\label{sec:gstates}
\begin{figure}[ht]
    \centering
    \includegraphics{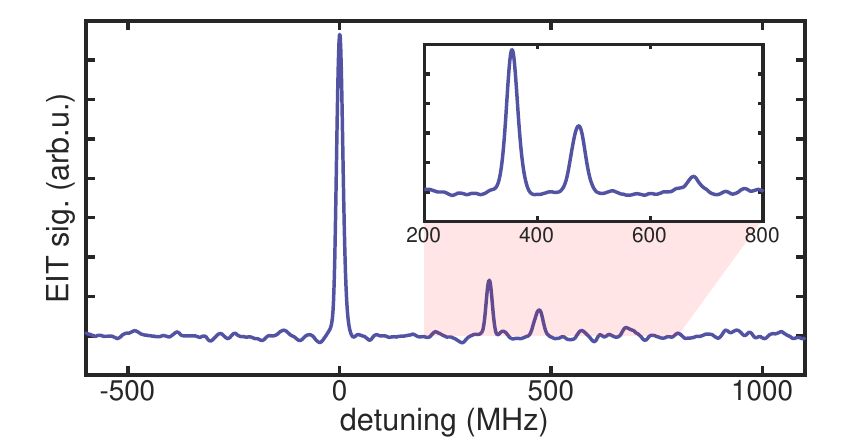}
    \caption{EIT spectrum of the possible $13g$ state at $49052.771\,$cm$^{-1}$. A fine-structure splitting of four resonances is visible, the inset shows a second measurement run confirming the fourth resonance feature.}
    \label{fig:gstate}
\end{figure}

Due to the angular momentum coupling to the ionic core, a two-photon excitation of a Rydberg state with $\ell = 4$ ($ng$ state) is possible. Below the first ionization threshold, the highest $n_\text{eff, 11/2} = 15$. Due to its large angular momentum, the corresponding quantum defect is expected to be close to zero. We identified several candidates with very small $\mu_{11/2}$ and further eliminated lines too close to other $ns$ or $nd$ resonances, especially when close to already identified perturbers. One candidate meets all requirements and additionally shows a distinct fine structure pattern different from the previous investigated series, see Fig. \ref{fig:gstate}. Its energy of $49052.771\,$cm$^{-1}$ would be compatible to a $13g$ state connected to the second ionization threshold.

\subsection{States above first ionization threshold}
\label{sec:abovethreshold}
We observed several states above the first ionization threshold, which we assigned to the second ionization threshold. With an effective quantum defect $\mu_{11/2} \approx 0.78$ we would identify them as belonging to the $nd$ series of the $J = 11/2$ state. We assign effective principal quantum numbers $n_\text{eff, 11/2}$ and measure their width as the FWHM value, as shown in Table \ref{tab:above_threshold}. Interestingly their widths vary by more then two orders of magnitude, presumably caused by drastically different lifetimes. A more systematic survey of these states could improve the MQDT modelling by constraining the parameters, as discussed in Sec.~\ref{sec:mqdt}.

\begin{table}[ht]
\begin{ruledtabular}
    \centering
    \caption{States above the first ionization threshold, with their measured full width at half maximum. Only the first two states have significantly increased widths, while the other states show a linewidth comparable to the states below the threshold.}
    \label{tab:above_threshold}
    \begin{tabular}{lrrr}
    $n_\text{eff, 11/2}$ & $\delta_\text{eff, 11/2}$ & $E_\text{meas}$ (cm$^{-1}$) & FWHM (MHz) \\
    \hline\hline
    $17d$ & 0.77 & 49284.43(1) & 350 \\
    $18d$ & 0.80 & 49330.22(1) & 30 \\
    $18d$ & 0.76 & 49332.16(1) & 2 \\
    $19d$ & 0.80 & 49369.82(1) & 4 \\
    \end{tabular}
\end{ruledtabular}
\end{table}

\subsection{Isotope shift}
\label{sec:isotopes}
To demonstrate the flexibility of our method and to exploit the large number of isotopes with high abundance in erbium, we record the isotope shift of the $n = 27d$ multiplets for four bosonic isotopes, \ts{164}Er, \ts{166}Er, \ts{168}Er, \ts{170}Er, as well as for two hyperfine states of the fermionic \ts{167}Er isotope, see Table \ref{tab:isotopes}. This shows the versatility of our approach, and our ability to switch easily between the addressed isotopes and hyperfine states. For the energy of the first photon, we take the value from \cite{Frisch2013} to calculate the total energy.

\begin{table}[ht]
\begin{ruledtabular}
    \centering
    \caption{Total energy of the lowest state of the $27d$ multiplets for different isotopes, with the energy difference between the Rydberg state and the intermediate ($E_{i,Ry}$) and the ground state ($E_\text{tot}$).}
    \label{tab:isotopes}
    \begin{tabular}{lrrr}
    isotope & $E_{i,Ry}$ (\wn{}) & $E_\text{tot}$ (\wn{})\\
    \hline\hline
    164 & 24169.102(5) & 49112.429(5) \\
    166 & 24169.106(5) & 49112.404(5) \\
    168 & 24169.105(5) & 49112.375(5) \\
    170 & 24169.108(5) & 49112.350(5) \\
    167\tu{(17/2,19/2)} & 24169.031(5) & 49112.313(5) \\
    167\tu{(9/2, 11/2)} & 24169.178(5) & 49112.465(5) \\
    \end{tabular}
\end{ruledtabular}
\end{table}

\section{Summary and outlook}
We observe about 550 Rydberg states in erbium with unprecedented precision and with principal quantum numbers as high as $n = 140$. By controlling the light polarization and applying a magnetic field, we can resolve the splitting between Zeeman sublevels and are able to assign the total angular momentum $J$ to the observed Rydberg series and measure their $g_J$-factors. 

The number and precision of states allowed for an accurate determination
of the MQDT parameters for the $ns$-Rydberg series. We were not able
to unambiguously determine the MQDT parameters for the more complicated
$nd$-Rydberg series. Two restricted models of the series were
able to reproduce many of the features of the $nd$-series. However,
the g-factors for several of the states and the details of the
perturbations were not accurately reproduced. Future technical improvements will allow a higher absolute accuracy of the measured Rydberg energies as well as enabling a more systematic survey of the states between the two ionization thresholds, thus providing an improved basis for understanding and modelling of the $nd$-Rydberg series.

Our spectroscopic study marks a first step on creating a new toolbox for Rydberg physics. While our first survey concentrated on an excitation scheme using a $6s$-electron, our vision is to employ new schemes using inner-shell $4f$-electrons. Additionally to the already new possibilities offered by two-electron atoms like Strontium or Ytterbium, we expect dramatically new physics to be present and that the active submerged shell will affect fundamental properties of these systems.

\begin{acknowledgments}
We thank Pascal Naubereith and Klaus Wendt for providing us unpublished spectroscopy data from the Mainz LARISSA project. This work is financially supported through an ERC Consolidator Grant (RARE, No. 681432). A.\,T.~acknowledges support through the FWF Lise-Meitner Fellowship No.~M~2683-N36. H.E. acknowledges support from the Israel Council for Higher Education. F.R. was supported by the National Science Foundation under Grant No. 1804026-PHY.  C.H.G. received support from the U.S. Department of Energy, Office of Science, Basic Energy Sciences, under Award Number DE-SC0010545. We also acknowledge the Innsbruck Laser Core Facility, financed by the Austrian Federal  Ministry of Science, Research and Economy.
\end{acknowledgments}

\appendix

\section{Multichannel quantum defect equations (MQDT)}\label{sec:mqdtApp}

The formulas used to calculate the energies and g-factors of the states are
given below. They follow the notation in Ref.~\cite{RBS2018} and are
given with little discussion. For a fuller derivation and discussion,
see Ref.~\cite{AGL1996} or \cite{RBS2018}.
The mass of $^{166}$Er$^+$ was taken from \cite{ErMass2000} 
to be $M_{166+}=165.930\; 293\; 1 u -m_e$ with
$u=1.660\; 539\; 066\; 60\times 10^{-27}$~kg and
$m_e=9.109\; 383\; 701\; 5\times 10^{-31}$~kg taken from the CODATA values.
The Rydberg constant, $R_{166}$, was taken to be scaled from the
CODATA $R_\infty$ value as
$R_{166}=$ 109 737.315 681 60~cm$^{-1}$
$\times M_{166+}/(M_{166+}+m_e)$.

\subsection{Calculated energies}\label{SecCalcE}

When the Rydberg electron is outside of the ionic core, the unphysical wave functions
can be written as
\begin{equation}
|\psi_{i}\rangle=\sum_{i'}|\Phi_{i'}\rangle [f_{i'}(r)\delta_{i',i} - g_{i'}(r) K_{i',i}]
\end{equation}
where the $f,g$ are the energy normalized, radial Coulomb functions
which are regular,irregular at the origin
and ${\bf K}$ is the real, symmetric K-matrix. See Ref.~\cite{AGL1996}
for the properties of these functions that depend on the radial position
of the Rydberg electron. The core states $|\Phi_{i'}\rangle $
contain all other degrees of freedom. For the Rydberg states described
above, the core states can be written as
\begin{equation}\label{EqPhiRyd}
|\Phi_{i}\rangle = |(J_{c,i} (s\ell_i)j_i)J_iM_i\rangle
\end{equation}
where $J_{c,i}$ is the total angular momentum of the core, $s$ is the
spin of the Rydberg electron, $\ell_i$ is the orbital angular momentum of
the Rydberg electron, $j_i$ is the total angular momentum of the
Rydberg electron, $J_i$ is the total angular momentum, and
$M_i$ is the related azimuthal quantum number. The order of parenthesis
is meant to indicate the order that the angular momenta are coupled
together.

The $|\psi_i\rangle $ function is
unphysical because the $f_{i'},g_{i'}$ functions diverge at large $r$ for closed channels
defined by $E<E_{c,i'}$.
At bound state energies, the $|\psi_{i}\rangle$ can be superposed
to give a physical eigen-function which converges to 0 at large $r$.
For the bound state at energy $E_b$, the superposition can be
written as
\begin{equation}|\psi_b\rangle = \sum_i |\psi_i\rangle \frac{\cos (
\beta_i )}{\nu_i^{3/2}}A_{i,b}
\end{equation}
and
the condition that determines the bound states is~\cite{AGL1996}
\begin{equation}\label{eqAdef}
\sum_{i} [\tan (\beta_{i'})\delta_{i',i}+K_{i',i}] \frac{\cos (\beta_{i})}{\nu_{i}^{3/2}}A_{i,b} = 0
\end{equation}
where $\beta_i=\pi (\nu_i-\ell_i)$ and the effective quantum number
is defined as $E=E_{c,i}-R_{166}/(2\nu_i^2)$ in terms of the total
energy, $E$, and the energy of the $i$th core state, $E_{c,i}$
with both energies in cm$^{-1}$.
The Rydberg constant is given in
App.~\ref{sec:mqdtApp} above. This condition only holds when
the term in $[\; ]$ has determinant equal to 0.
The normalization condition when the $K$-matrix
is slowly varying with energy is
\begin{equation}
\sum_i A^2_{i,b} = 1
\end{equation}
which, with Eq.~(\ref{eqAdef}), defines the $A_{i,b}$ within an
irrelevant, overall sign.

\subsection{Calculated g-factor}\label{Secgfac}

The g-factor for individual states can be approximately calculated from
the MQDT parameters by assuming the contribution is negligible 
when the Rydberg electron is within the ionic core region. This situation
is covered by Eq.~(4.1.2) of Ref.~\cite{AGL1996} (or similarly
Eqs.~(15) and (16) of Ref.~\cite{RBS2018}):
\begin{equation}
g_b=\sum_{i,i'}A_{i,b}\langle\Phi_i|\hat{g}|\Phi_{i'}\rangle O_{ib,i'b}A_{i'b}
\end{equation}
where the overlap matrix is
\begin{equation}
O_{ib,i'b}=
\frac{2\sqrt{\nu_{ib}\nu_{i'b}}}{\nu_{ib}+\nu_{i'b}}
\frac{\sin  (\beta_{ib}-\beta_{i'b})}{(\beta_{ib}-\beta_{i'b})}
\end{equation}
with $O_{ib,ib}=1$.
The $g$-operator matrix element is complicated but only uses
Eqs.~(3.7.9),
(5.4.1), (5.4.3), (7.1.7) and (7.1.8)
of Ref.~\cite{ARE1974}. The matrix element is
\begin{eqnarray}
\langle\Phi_i|\hat{g}|\Phi_{i'}\rangle &=&\frac{1}{M_i}
\langle\Phi_i|g_{c,i}J_{c,z}+\ell_z+g_s s_z|\Phi_{i'}\rangle\nonumber \\
&=&\frac{\langle\Phi_i||g_{c,i}J_{c}^{(1)}+\ell^{(1)}+g_s s^{(1)}||\Phi_{i'}\rangle}{\Lambda (J_i)}
\end{eqnarray}
where $g_c = 1.230$ for the $J_c=13/2$ state and 1.101 for the
$J_c=11/2$ state \cite{NIST_ASD,McNally1959} and $g_s=2.002319...$.
The $\Lambda (x)\equiv \sqrt{(2x+1)(x+1)x}$.
The reduced matrix elements when $\ell$ is the
same for $i$ and $i'$ are
\begin{eqnarray}
\langle\Phi_i||\ell^{(1)}||\Phi_{i'}\rangle &=&
\delta_{J_{c,i}J_{c,i'}}{\cal G}_1{\cal G}_2\Lambda (\ell )
\nonumber\\
\langle\Phi_i||s^{(1)}||\Phi_{i'}\rangle &=&
\delta_{J_{c,i}J_{c,i'}}{\cal G}_1{\cal G}_3\Lambda (s )
\nonumber\\
\langle\Phi_i||J_c^{(1)}||\Phi_{i'}\rangle &=&
\delta_{j_ij_{i'}}\delta_{J_{c,i}J_{c,i'}}{\cal G}_4\Lambda (J_{c,i} )
\end{eqnarray}
where we have made the approximation that the core angular momentum
operator does not mix core states with different $J_c$ and
\begin{eqnarray}
{\cal G}_1&=&(-1)^{J_{c,i}+j_{i'}+J+1}
(2J+1)
\begin{Bmatrix}
j_i&J&J_{c,i}\\
J&j_{i'}&1
\end{Bmatrix}
\nonumber\\
{\cal G}_2&=&(-1)^{s+\ell+j_i+1}
[j_i,j_{i'}]
\begin{Bmatrix}
\ell &j_i&s\\
j_{i'}&\ell &1
\end{Bmatrix}
\nonumber\\
{\cal G}_3&=&(-1)^{s+\ell+j_{i'}+1}
[j_i,j_{i'}]
\begin{Bmatrix}
s &j_i&\ell\\
j_{i'}&s &1
\end{Bmatrix}
\nonumber\\
{\cal G}_4&=&(-1)^{J_{c,i}+j_{i}+J+1}
(2J+1)
\begin{Bmatrix}
J_{c,i} &J&j_i\\
J&J_{c,i'} &1
\end{Bmatrix}
\end{eqnarray}

\subsection{Frame transformation approximation}\label{SecFT}

The energy dependent $K$-matrix exactly determines the bound state
energies and approximately allows evaluation of other state properties
(e.g. the g-factor). Unfortunately, the electronic structure of
Er is too complicated for an accurate, ab initio calculation. 
Nevertheless, there are two possible paths to obtain the $K$-matrix.
The first is to use the experimental data, without any guidance from
a model, to obtain
a fit of the $K$-matrix. This is described in Sec.~\ref{Secns}. The other
is to utilize an approximation to restrict the possible values of the
$K$-matrix before using the experimental energies to fit the
elements of the $K$-matrix. This section describes this second method
where the approximation is based on a frame transformation. The
following section will use a different physical idea to reduce
the number of parameters defining the $K$-matrix.

The frame transformation approximation assumes there is a channel
coupling that diagonalizes the $K$-matrix when the electron is
in the core region. The Rydberg states are attached to fine structure
split core states of Er$^+$: $4f^{12}(^3H_6)6s_{1/2}J_c$ with $J_c=13/2$
and $11/2$ being the ground and first excited state, respectively.
The idea is to call the angular momentum of the $4f^{12}$ inner electrons
$J_f=6$ and the angular momentum of the valence $6s$ electron
$J_s=1/2$. In terms of the notation of App.~\ref{SecCalcE}, the
channel states when the electron is outside of the core, Eq.~(\ref{EqPhiRyd}),
is expanded to
\begin{equation}
|\Phi_i\rangle = |((J_fJ_s)J_{c,i}(s\ell_i)j_i)J_iM_i\rangle .
\end{equation}
We assume that the coupling that leads to a diagonal $K$-matrix is given
by
\begin{equation}
|\Phi^{in}_{i'}\rangle =|(J_f((J_ss)S_{o,i'}\ell_{i'})J_{o,i'})J_{i'}M_{i'}\rangle
\end{equation}
where the order of coupling is: the spin of the $6s$ and Rydberg electron
coupled to give total outer spin, $S_{o,i}$, then the total outer spin
is coupled to the
orbital angular momentum of the Rydberg electron, $\ell_i$, to give
the total outer angular momentum, $J_{o,i}$,
then the angular momentum of the inner
$4f^{12}$, $J_f=6$ is coupled to the total outer angular momentum
to give the total angular momentum. The overlap matrix between these
couplings are only nonzero if $J_i=J_{i'}$, $M_i=M_{i'}$, and
$\ell_i=\ell_{i'}$. For Er, the $s=1/2$ and is automatically the same.

With the help of the intermediate coupling $|(J_f(J_s(s\ell )j)J_o)JM\rangle $,
the projection of the two coupling schemes is derived from
Eq.~(6.1.5) of Ref.~\cite{ARE1974} to give
\begin{eqnarray}
\langle\Phi_i|\Phi^{in}_{i'}\rangle&=&
[j_i,S_{o,i'},J_{c,i},J_{o,i'}](-1)^{J_f+2J_s+j_i+J+s+\ell+J_{o,i'}}
\nonumber\\
&\null &
\begin{Bmatrix}
J_f &J_s&J_{c,i}\\
j_i&J &J_{o,i'}
\end{Bmatrix}
\begin{Bmatrix}
J_s &s&S_{o,i'}\\
\ell&J_{o,i'} &j_{i}
\end{Bmatrix}
\end{eqnarray}
with $[j_1,j_2...]=\sqrt{(2j_1+1)(2j_2+1)...}$.

For the case where the Rydberg electron has $s$ character, there are 2
channels for $J=6$ and one channel for $J=7$. Thus, the frame
transformation is only useful for $J=6$. The two $|\Phi\rangle$
states can be defined as $|J_c\rangle\equiv ((^3H_66s_{1/2})J_c ns_{1/2})
J=6$ with $J_c=13/2$ and $11/2$ while the two
$|\Phi^{in}\rangle$ can be defined as $|J_o\rangle\equiv
(^3H_6 (6s_{1/2}ns_{1/2})^{2S_o+1}S_{S_o})J=6$ with $S_o=0$ and
1. In this simplified notation, $\langle 13/2|0\rangle =
\langle 11/2|1\rangle =\sqrt{7/13}$ and
$\langle 13/2|1\rangle =
-\langle 11/2|0\rangle =\sqrt{6/13}$.

For the case where the Rydberg electron has $d$ character, there
are 4 channels for $J=6$ and 7 and 3 channels for $J=8$. The
$|\Phi\rangle$ can be defined as $|J_c,j\rangle\equiv
((^3H_66s_{1/2})J_c nd_{j})J$. Both $J=6$ and 7 have the 4 couplings
$|13/2,3/2\rangle$, $|13/2,5/2\rangle$, $|11/2,3/2\rangle$, and
$|11/2,5/2\rangle$ while the $J=8$ only has
$|13/2,3/2\rangle$, $|13/2,5/2\rangle$,  and
$|11/2,5/2\rangle$. The 
$|\Phi^{in}\rangle$ can be defined as $|S_o,J_o\rangle\equiv 
(^3H_6 (6s_{1/2}nd_{1/2})^{2S_o+1}D_{J_o})J$.
Both $J=6$ and 7 have the 4 couplings, $|0,2\rangle$, $|1,1\rangle$,
$|1,2\rangle$, and $|1,3\rangle$ while the $J=8$ only has
$|0,2\rangle$, $|1,2\rangle$, and $|1,3\rangle$. If we let the
quantum defects vary with $J$, then there are 11 free parameters
for all of the $K$-matrices.

\subsection{Interaction inspired K-matrix approximation}\label{SecQP}

A completely different method for parameterizing the $K$-matrix
for the $nd$ channels is
to identify interactions that lead to coupling between channels
or shifts in channels. This section discusses some of the important
interactions.

There are two shifts that are expected.
The first is the average $K$-matrix will depend on $J$. The
second is the shift due to the spin-orbit interaction. These
lead to terms of the form:
\begin{equation}\label{EqKdiag}
K^{(d)}_{ii'}=\delta_{i,i'}[\tan (\pi\mu_{J}) + \delta_{j_i,3/2}\tan (\pi\mu_{3/2})]
\end{equation}
where we have put all of the spin-orbit shift into the $nd_{3/2}$
channels.

Another important interaction arises from a second rank coupling
of the core state interacting with the $nd$ electron. This type of
interaction can arise through a quadrupole moment of the ionic
core or from an anisotropic polarizability of the core \cite{Becher2018pra}.  Other long range interaction terms between the Rydberg electron and the anisotropic ionic core could be introduced, such as the vector hyperpolarizabity term, but the present level of theory has not yet taken such interactions into account \cite{ClarkGreene1999rmp,Watanabe1980pra}.
The angular part of the matrix element arises from
\begin{equation}
Q_{ii'}=\langle\Phi_i |P_2(\hat{r}_c\cdot\hat{r})|\Phi_{i'}\rangle
\end{equation}
where $|\Phi_i\rangle$ are from Eq.~(\ref{EqPhiRyd}) and the
$P_2(\hat{r}_c\cdot\hat{r})$ is the Legendre polynomial of
the Rydberg electron dotted into core electrons. To evaluate
this, we use the fact that the $4f^{12}$ electrons are coupled
as $S_f=1$, $L_f=5$, and $J_f=6$ before coupling to the
$6s$ electron to give $J_c$.
Using Eqs.~(5.4.6), (7.1.6), (7.1.7),
and (7.1.8) of Ref.~\cite{ARE1974}, we find $Q_{ii'}=C\; q_{ii'}$
\begin{eqnarray}
q_{ii'}&=&(-1)^{2J_{c,i'}+2j_i+J }\frac{[j_i,j_{i'},J_{c,i},J_{c,i'}]}{[J]}
\begin{Bmatrix}
\ell &j_i&s\\
j_{i'}&\ell &2
\end{Bmatrix}
\nonumber\\
&\null &\times
\begin{Bmatrix}
J &j_i&J_{c,i}\\
2&J_{c,i'} &j_{i'}
\end{Bmatrix}
\begin{Bmatrix}
J_f &J_{c,i}&J_s\\
J_{c,i'}&J_f &2
\end{Bmatrix}
\end{eqnarray}
when making the approximation $\ell_i=\ell_{i'}.$
Since the size of the second rank coupling is unknown, we add this to
the $K$-matrix with a fitting parameter to obtain the total:
\begin{equation}
K_{ii'}=K^{(d)}_{ii'} + C_q q_{ii'}
\end{equation}

If we stop at this level, there are 5 free parameters in the $K$-matrix.

\section{Estimation of the spectral pattern of different {\it J} states of the {\it nd} manifold}\label{Appendix:Specsim}

\begin{figure}[ht]
    \centering
    \includegraphics{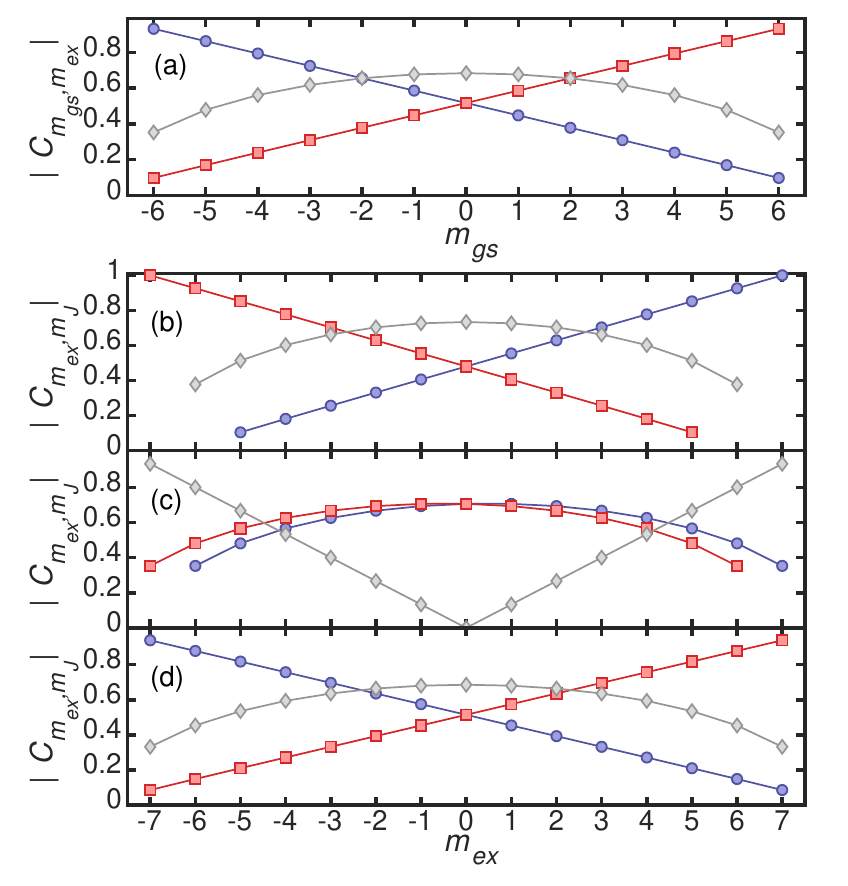}
    \caption{Absolute values of the transition matrix elements for the first transition a) $|J_{gs}=6,m_{gs}\rangle\rightarrow|J_{ex}=7,m_{ex}\rangle$ as well as for the three possibilities of the second transition b) $|J_{ex}=7,m_{ex}\rangle\rightarrow|J=6,m_J\rangle$, c) $|J_{ex}=7,m_{ex}\rangle\rightarrow|J=7,m_J\rangle$ and d) $|J_{ex}=7,m_{ex}\rangle\rightarrow|J=8,m_J\rangle$. Circles (squares) encode $\sigma^-$ ($\sigma^+$) transitions, diamonds display $\pi$ transitions.}
    \label{fig:CGC}
\end{figure}

For a simple estimation of the relative EIT resonance strength we create a list of all dipole allowed transitions from the groundstate $|J_{gs}=6,m_{gs}\rangle$ over the intermediate state $|J_{ex}=7,m_{ex}\rangle$ to the Rydberg states $|J=6,7,8,m_J\rangle$. We calculate the transition matrix elements for each single photon transition using the Wigner $3j$-symbol, leading to
\begin{eqnarray}
C_{m_{gs},m_{ex}} & = & (-1)^{J_{gs}-1+m_{ex}}\sqrt{2J_{gs}+1}
\begin{pmatrix}
J_{gs} & 1 & J_{ex}\\
m_{gs} & \Delta m & -m_{ex}\nonumber
\end{pmatrix}
\end{eqnarray}
for the first transition and
\begin{eqnarray}
C_{m_{ex},m_J} & = & (-1)^{J_{ex}-1+m_J}\sqrt{2J_{ex}+1}
\begin{pmatrix}
J_{ex} & 1 & J\\
m_{ex} & \Delta m & -m_J\nonumber
\end{pmatrix}
\end{eqnarray}
for the transition from the intermediate to the Rydberg state. Figure \ref{fig:CGC} shows the calculated values as a function of the $m_J$ state for all possible $J$ values.

\begin{figure}[ht]
    \centering
    \includegraphics{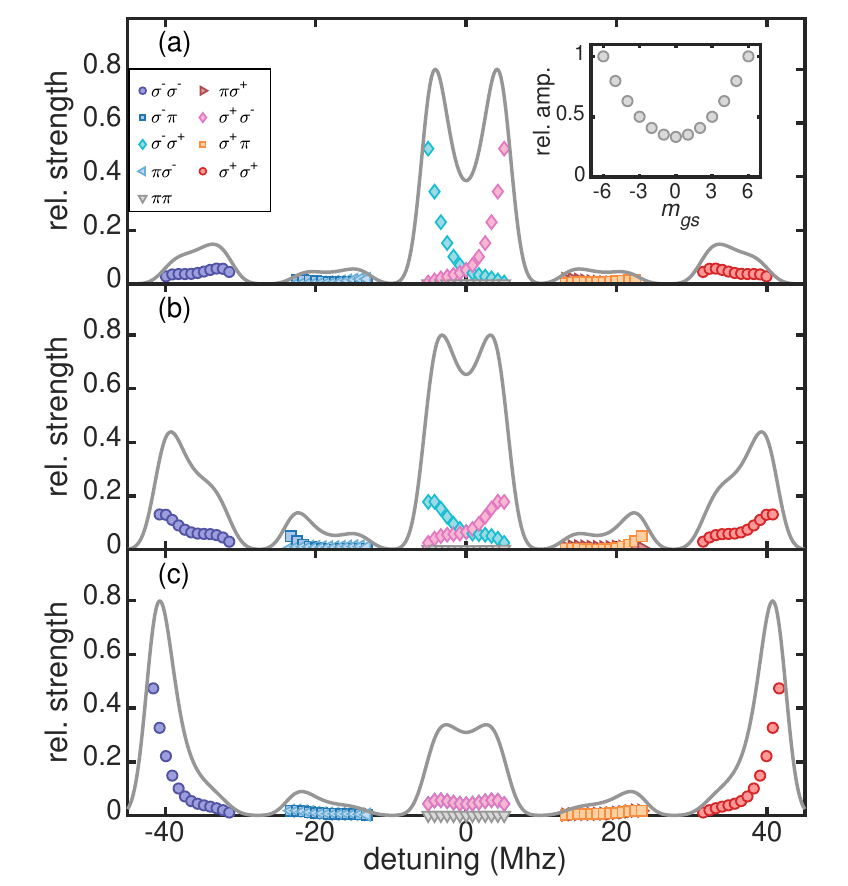}
    \caption{Estimation of the spectral pattern given the relative 2-photon transition strength for a) $|J=6,m_J\rangle$, b) $|J=7,m_J\rangle$ and c) $|J=8,m_J\rangle$. Different colors and symbols encode the transition type (see legend). The inset shows the assumed relative distribution of populations among the groundstate manifold.}
    \label{fig:transstrength}
\end{figure}

For each possible combination we can calculate the relative 2-photon strength $|C_{m_{gs},m_{ex}}\times C_{m_{ex},m_J}|$ and multiply it with our estimated amplitudes of our light polarizations (here we assume $5\%\,\,\pi$-light) and the corresponding amplitude of the population of the Zeeman level of the groundstate. Here, the population amplitudes account for the effects of optical pumping during the spectroscopy. Now we can plot it as a function of the 2-photon detuning for a given magnetic field, in our case $10.7\,$G, see Figure \ref{fig:transstrength}. Here we exemplary assume a g-factor $g_J = 1.22$ for the Rydberg state. The solid line gives a total sum of all transitions, taking a gaussian lineshape with a width of $2\,$MHz and the corresponding 2-photon strength as amplitude for each transition. 

Each $J$ has its own characteristic spectral pattern: For $J=6$ a strong central peak originating from $\Delta m_J=0$ transitions (splitted for large $g_J$) with two weaker side peaks ($\Delta m_J=\pm 2$ transitions) is expected, while for $J=8$ the central peak should be much weaker and the side peaks should be strongest. For $J=7$ the central peak as well as the side peaks have roughly the same strength, and only here also the $\Delta m_J=\pm 1$ transitions have a significant strength and might be visible.

\section{Estimation of the experimental uncertainties and frequency inaccuracies}\label{Appendix:Error}

The used wavemeter has an absolute frequency accuracy of \unit \pm60 MHz, given as $3\sigma$ interval \cite{Toptica2021}. This error is partially systematic, depending on various environmental conditions. Long-time measurements on a laser locked to an atomic transition in our lab shows frequency deviations of typical \unit 20 MHz within one hour. We therefore assume a fundamental $1\sigma$ error of \unit \pm20 MHz for single measurements and an additional $1\sigma$ systematic uncertainty of \unit \pm10 MHz when combining multiple measurement sets. 

To be able to give the absolute energy of the measured states we have to add the first transition, whose energy we also measure with the same wavemeter. We always measure the fundamental frequency of both, probe and coupling beam, before frequency doubling. We estimate our wavemeter-limited accuracy for single state energies therefore to about \unit \pm69 MHz i.e. \wn{$\pm0.0023$}. The final error in the determination of the ionization threshold includes the error of the fit itself and the combined systematic uncertainty of probe- and coupling frequency (\wn{$\pm0.0013$}), resulting in an estimated accuracy of \wn{$\pm0.0023$}.

An additional uncertainty in our resonance position energies results from our wavelength monitoring. While scanning the coupling frequency using an analog voltage ramp, we continuously read out the measured frequencies from the wavemeter. Due to the limited and variable read-out speed (depending on the light intensity coupled to the wavemeter), which is not synchronized to the analog ramp, we record frequency traces during several scans and take the maximum and minimum measured wavelength to calibrate the frequency axis of the scans.

For most of the wide scans used in our survey we estimate the absolute uncertainty of the resonance positions with \wn{0.07}, mainly limited by nonlinearities within each frequency scan. For the $nd$ states investigated for the g-factor, the smaller scan ranges minimize those nonlinearities and allows us to reduce the absolute uncertainty to \wn{0.005}, now primarily limited by the absolute accuracy of the wavemeter.

The main uncertainty in the fitting of the g-factor is the determination of the proper frequency axis scale. For these very narrow scan ranges, the finite read-out speed of the wavemeter results in significant variations of the frequency axis scale of up to $20\%$ when comparing different experimental runs where the scan range should have been identical. We account for this by scaling all frequency axes of one set to the maximum range, as the described frequency axis calibration will always lead to an underestimation of the scan range. The given error of the g-factor takes into account the fit error itself, the uncertainty of the $B$-field calibration and the rescaling factor.

\clearpage
\onecolumngrid
\section{List of all observed states}\label{sec:allstates}

\begin{ruledtabular}
\gdef\tmphead{\strut\rlap{$n_\text{eff, 13/2}$} & \strut \rlap{$E (\wn{})$}\\}

\begin{longtable*}{lllllllllllll}
\caption{Table listing all observed features, listed by effective quantum number $n_\text{eff}$, and term energy. Errors in energy are \wn{0.07}. Some features have been measured with higher precision, see main text.
\label{tab:all_states}}
\\ \hline\hline
\rlap{$n_\text{eff, 13/2}$} & \strut \rlap{$E (\wn{})$}\\
\hline
\endfirsthead
\hline\hline
\tmphead
\hline
\endhead
\hline\hline
\endfoot
15 & 48828.76 \\ 
16 & 48841.30 & 48842.77 & 48842.83 & 48843.43 & 48844.77 & 48845.23 & 48863.11 & 48863.98 \\ 
17 & 48888.33 & 48889.26 & 48889.80 & 48889.93 & 48891.47 & 48892.07 & 48892.73 & 48909.31 & 48909.33 \\ 
18 & 48929.32 & 48929.99 & 48930.00 & 48930.01 & 48930.03 & 48930.87 & 48931.46 & 48931.71 & 48931.94 & 48945.19 \\ 
19 & 48963.98 & 48964.28 & 48964.32 & 48964.92 & 48964.98 & 48965.52 & 48968.26 & 48969.00 & 48979.36 \\ 
20 & 48992.82 & 48993.07 & 48993.20 & 48993.44 & 48993.70 & 49002.52 \\ 
21 & 49015.08 & 49016.30 & 49016.42 & 49017.03 & 49017.42 & 49026.46 & 49028.00 \\ 
22 & 49038.45 & 49038.83 & 49039.17 & 49039.37 & 49039.43 & 49047.31 & 49052.80 & 49052.81 & 49052.81 & 49052.82 \\ 
23 & 49056.91 & 49056.93 & 49057.10 & 49057.37 & 49057.79 & 49058.02 & 49058.13 & 49064.84 \\ 
24 & 49073.44 & 49073.97 & 49073.99 & 49074.33 & 49074.39 & 49074.79 & 49080.13 & 49081.41 \\ 
25 & 49087.98 & 49088.54 & 49088.66 & 49088.68 & 49088.83 & 49093.78 \\ 
26 & 49100.58 & 49100.86 & 49101.22 & 49101.51 & 49101.54 & 49101.62 & 49101.70 & 49101.72 & 49105.45 \\ 
27 & 49111.45 & 49112.39 & 49112.56 & 49112.82 & 49112.98 & 49113.06 & 49113.14 & 49117.95 \\ 
28 & 49122.73 & 49122.76 & 49122.86 & 49123.23 & 49123.26 & 49123.29 & 49127.35 \\ 
29 & 49132.01 & 49132.16 & 49132.39 & 49132.49 & 49132.53 & 49132.63 & 49136.08 \\ 
30 & 49140.44 & 49140.60 & 49140.75 & 49140.96 & 49140.98 & 49144.04 \\ 
31 & 49147.99 & 49148.12 & 49148.28 & 49148.38 & 49148.46 & 49148.50 & 49151.25 & 49151.27 \\ 
32 & 49154.90 & 49155.02 & 49155.17 & 49155.25 & 49155.35 & 49157.12 & 49157.74 & 49157.82 \\ 
33 & 49160.85 & 49161.25 & 49161.41 & 49161.45 & 49161.52 & 49161.60 & 49163.83 \\ 
34 & 49167.02 & 49167.12 & 49167.15 & 49167.22 & 49167.29 & 49169.25 \\ 
35 & 49172.18 & 49172.38 & 49172.45 & 49172.51 & 49174.33 \\ 
36 & 49177.01 & 49177.19 & 49177.25 & 49177.28 & 49177.31 & 49178.94 \\ 
37 & 49181.43 & 49181.51 & 49181.60 & 49181.63 & 49181.69 & 49181.73 \\ 
38 & 49185.54 & 49185.70 & 49185.75 & 49185.80 & 49186.73 \\ 
39 & 49188.90 & 49188.91 & 49189.34 & 49189.52 & 49189.54 & 49189.55 & 49189.57 & 49191.35 \\ 
40 & 49192.84 & 49192.98 & 49193.02 & 49193.04 & 49193.07 & 49194.52 \\ 
41 & 49196.09 & 49196.22 & 49196.26 & 49196.28 & 49196.31 & 49197.61 \\ 
42 & 49199.13 & 49199.25 & 49199.29 & 49199.33 & 49200.52 \\ 
43 & 49201.95 & 49202.08 & 49202.11 & 49202.15 \\ 

45 & 49207.04 & 49207.08 & 49207.14 & 49207.16 & 49207.20 & 49208.14 \\ 
46 & 49209.35 & 49209.42 & 49209.45 & 49209.49 & 49210.35 \\ 
47 & 49211.49 & 49211.53 & 49211.58 & 49211.61 & 49211.64 & 49211.69 & 49211.69 & 49212.43 & 49213.08 \\ 
48 & 49213.51 & 49213.53 & 49213.59 & 49213.61 & 49213.62 & 49213.64 & 49214.30 \\ 

50 & 49217.20 & 49217.28 & 49217.29 & 49217.32 & 49217.32 & 49218.53 \\ 
51 & 49218.85 & 49218.95 & 49218.96 & 49219.01 & 49219.02 & 49219.63 \\ 
52 & 49220.48 & 49220.55 & 49220.57 & 49220.59 & 49221.17 \\ 
53 & 49221.99 & 49222.04 & 49222.07 & 49222.08 & 49222.64 \\ 
54 & 49223.40 & 49223.45 & 49223.47 & 49223.49 \\ 
55 & 49224.73 & 49224.79 & 49224.81 & 49224.82 & 49225.34 \\ 
56 & 49226.00 & 49226.00 & 49226.06 & 49226.08 & 49226.73 \\ 
57 & 49227.23 & 49227.27 & 49227.29 & 49227.30 & 49227.75 \\ 
58 & 49228.37 & 49228.40 & 49228.42 & 49228.45 & 49228.45 & 49228.83 & 49228.86 & 49229.11 & 49229.20 \\ 
59 & 49229.45 & 49229.47 & 49229.49 & 49229.51 & 49229.52 & 49229.83 & 49229.87 \\ 
60 & 49230.49 & 49230.53 & 49230.54 & 49230.55 \\ 
61 & 49231.29 & 49231.35 & 49231.38 & 49231.40 & 49231.89 \\ 
62 & 49232.59 & 49232.64 & 49232.67 & 49232.67 & 49232.69 \\ 

64 & 49234.13 & 49234.16 & 49234.17 & 49234.18 & 49234.60 & 49234.62 & 49234.62 & 49234.64 & 49234.76 \\ 
65 & 49234.93 & 49234.97 & 49234.98 & 49234.99 & 49235.29 \\ 
66 & 49235.72 & 49235.74 & 49235.76 & 49235.77 & 49235.78 & 49236.04 & 49236.10 & 49236.14 & 49236.17 \\ 
67 & 49236.50 \\ 
68 & 49237.05 & 49237.14 & 49237.17 & 49237.18 & 49237.19 & 49237.46 \\ 
69 & 49237.83 & 49237.86 & 49237.87 & 49237.87 & 49237.89 & 49238.13 \\ 

71 & 49239.10 & 49239.14 & 49239.15 \\ 
72 & 49239.69 & 49239.70 & 49239.71 & 49239.71 & 49239.71 & 49239.72 & 49239.94 \\ 
73 & 49240.26 & 49240.27 & 49240.28 & 49240.29 \\

82 & 49244.51 & 49244.53 & 49244.54 & 49244.55 \\ 
83 & 49244.90 & 49244.91 & 49244.92 \\ 
84 & 49245.28 & 49245.28 & 49245.29 & 49245.31 & 49245.32 \\ 
85 & 49245.64 & 49245.65 & 49245.66 & 49245.67 & 49245.68 \\ 
86 & 49245.98 & 49245.98 & 49245.99 & 49246.00 & 49246.00 \\

92 & 49247.86 & 49247.87 \\ 
93 & 49248.11 & 49248.13 & 49248.14 & 49248.16 \\ 
94 & 49248.39 & 49248.39 & 49248.41 \\ 
95 & 49248.66 & 49248.67 & 49248.67 & 49248.68 \\ 
96 & 49248.90 & 49248.91 & 49248.92 & 49248.93 \\ 
97 & 49249.15 & 49249.15 & 49249.17 & 49249.17 & 49249.17 & 49249.19 \\ 
98 & 49249.38 & 49249.39 & 49249.45 & 49249.47 \\ 
99 & 49249.61 & 49249.62 & 49249.63 & 49249.65 \\ 
100 & 49249.85 & 49249.85 & 49249.86 & 49249.86 & 49249.93 & 49249.95 \\ 
101 & 49250.02 & 49250.03 & 49250.05 & 49250.06 & 49250.16 & 49250.18 \\ 
102 & 49250.24 & 49250.25 & 49250.25 & 49250.26 \\ 
103 & 49250.48 & 49250.49 \\ 
104 & 49250.65 & 49250.67 \\ 
105 & 49250.85 & 49250.86 \\ 
106 & 49251.05 & 49251.08 \\ 
107 & 49251.20 & 49251.21 \\ 
108 & 49251.43 & 49251.44 \\ 
109 & 49251.55 & 49251.57 & 49251.59 \\ 
110 & 49251.73 & 49251.74 & 49251.76 \\ 
111 & 49251.89 & 49251.90 & 49251.91 & 49251.91 & 49251.93 & 49251.93 \\ 
112 & 49252.04 & 49252.05 & 49252.07 & 49252.08 \\ 
113 & 49252.19 & 49252.19 & 49252.22 & 49252.23 \\ 
114 & 49252.34 & 49252.35 & 49252.36 \\ 
115 & 49252.45 & 49252.46 & 49252.46 & 49252.47 \\

118 & 49252.92 & 49252.95 & 49252.96 \\ 
119 & 49253.04 & 49253.08 & 49253.08 \\ 
120 & 49253.16 & 49253.19 & 49253.20 \\ 
121 & 49253.27 & 49253.30 & 49253.31 \\ 
122 & 49253.37 & 49253.41 & 49253.42 & 49253.45 & 49253.46 & 49253.48 \\ 
123 & 49253.51 & 49253.60 \\ 
124 & 49253.65 & 49253.66 \\ 
125 & 49253.72 & 49253.76 & 49253.77 & 49253.83 \\ 
126 & 49253.87 & 49253.87 & 49253.93 \\ 
127 & 49253.97 & 49253.97 & 49254.02 & 49254.04 \\ 
128 & 49254.09 & 49254.09 & 49254.15 \\ 
129 & 49254.19 & 49254.19 & 49254.24 \\ 
130 & 49254.28 & 49254.28 & 49254.29 & 49254.31 & 49254.33 & 49254.35 \\ 
131 & 49254.38 & 49254.38 & 49254.39 & 49254.40 & 49254.43 \\ 
132 & 49254.44 & 49254.49 & 49254.49 & 49254.52 & 49254.54 \\ 
133 & 49254.58 & 49254.62 \\ 
134 & 49254.66 & 49254.70 & 49254.71 \\ 
135 & 49254.73 & 49254.74 & 49254.77 & 49254.77 & 49254.78 & 49254.78 & 49254.78 & 49254.79 \\ 
136 & 49254.85 & 49254.86 & 49254.89 \\ 
137 & 49254.93 & 49254.97 \\ 
138 & 49255.01 & 49255.05 \\ 
139 & 49255.09 & 49255.13 \\ 
140 & 49255.15 & 49255.17 \\ 
\hline $n_\text{eff, 11/2}$ & $E$ (\wn{}) \\ \hline
11 & 48830.62 & 48830.66\\ 
12 & 48970.09\\ 
14 & 49157.12 & 49189.47\\ 
15 & 49219.46 & 49228.23 & 49229.09 & 49229.11\\ 
16 & 49284.46\\ 
17 & 49330.24 & 49332.19\\ 
18 & 49369.84 
\end{longtable*}
\end{ruledtabular}

\clearpage
\twocolumngrid


\begin{thebibliography}{55}%
\makeatletter
\providecommand \@ifxundefined [1]{%
 \@ifx{#1\undefined}
}%
\providecommand \@ifnum [1]{%
 \ifnum #1\expandafter \@firstoftwo
 \else \expandafter \@secondoftwo
 \fi
}%
\providecommand \@ifx [1]{%
 \ifx #1\expandafter \@firstoftwo
 \else \expandafter \@secondoftwo
 \fi
}%
\providecommand \natexlab [1]{#1}%
\providecommand \enquote  [1]{``#1''}%
\providecommand \bibnamefont  [1]{#1}%
\providecommand \bibfnamefont [1]{#1}%
\providecommand \citenamefont [1]{#1}%
\providecommand \href@noop [0]{\@secondoftwo}%
\providecommand \href [0]{\begingroup \@sanitize@url \@href}%
\providecommand \@href[1]{\@@startlink{#1}\@@href}%
\providecommand \@@href[1]{\endgroup#1\@@endlink}%
\providecommand \@sanitize@url [0]{\catcode `\\12\catcode `\$12\catcode
  `\&12\catcode `\#12\catcode `\^12\catcode `\_12\catcode `\%12\relax}%
\providecommand \@@startlink[1]{}%
\providecommand \@@endlink[0]{}%
\providecommand \url  [0]{\begingroup\@sanitize@url \@url }%
\providecommand \@url [1]{\endgroup\@href {#1}{\urlprefix }}%
\providecommand \urlprefix  [0]{URL }%
\providecommand \Eprint [0]{\href }%
\providecommand \doibase [0]{http://dx.doi.org/}%
\providecommand \selectlanguage [0]{\@gobble}%
\providecommand \bibinfo  [0]{\@secondoftwo}%
\providecommand \bibfield  [0]{\@secondoftwo}%
\providecommand \translation [1]{[#1]}%
\providecommand \BibitemOpen [0]{}%
\providecommand \bibitemStop [0]{}%
\providecommand \bibitemNoStop [0]{.\EOS\space}%
\providecommand \EOS [0]{\spacefactor3000\relax}%
\providecommand \BibitemShut  [1]{\csname bibitem#1\endcsname}%
\let\auto@bib@innerbib\@empty
\bibitem [{\citenamefont {Jaksch}\ \emph {et~al.}(2000)\citenamefont {Jaksch},
  \citenamefont {Cirac}, \citenamefont {Zoller}, \citenamefont {Rolston},
  \citenamefont {C\^ot\'e},\ and\ \citenamefont {Lukin}}]{Jaksch2000}%
  \BibitemOpen
  \bibfield  {author} {\bibinfo {author} {\bibfnamefont {D.}~\bibnamefont
  {Jaksch}}, \bibinfo {author} {\bibfnamefont {J.~I.}\ \bibnamefont {Cirac}},
  \bibinfo {author} {\bibfnamefont {P.}~\bibnamefont {Zoller}}, \bibinfo
  {author} {\bibfnamefont {S.~L.}\ \bibnamefont {Rolston}}, \bibinfo {author}
  {\bibfnamefont {R.}~\bibnamefont {C\^ot\'e}}, \ and\ \bibinfo {author}
  {\bibfnamefont {M.~D.}\ \bibnamefont {Lukin}},\ }\href {\doibase
  10.1103/PhysRevLett.85.2208} {\bibfield  {journal} {\bibinfo  {journal}
  {Phys. Rev. Lett.}\ }\textbf {\bibinfo {volume} {85}},\ \bibinfo {pages}
  {2208} (\bibinfo {year} {2000})}\BibitemShut {NoStop}%
\bibitem [{\citenamefont {Brennen}\ \emph {et~al.}(2000)\citenamefont
  {Brennen}, \citenamefont {Deutsch},\ and\ \citenamefont
  {Jessen}}]{Brennen2000}%
  \BibitemOpen
  \bibfield  {author} {\bibinfo {author} {\bibfnamefont {G.~K.}\ \bibnamefont
  {Brennen}}, \bibinfo {author} {\bibfnamefont {I.~H.}\ \bibnamefont
  {Deutsch}}, \ and\ \bibinfo {author} {\bibfnamefont {P.~S.}\ \bibnamefont
  {Jessen}},\ }\href {\doibase 10.1103/PhysRevA.61.062309} {\bibfield
  {journal} {\bibinfo  {journal} {Phys. Rev. A}\ }\textbf {\bibinfo {volume}
  {61}},\ \bibinfo {pages} {062309} (\bibinfo {year} {2000})}\BibitemShut
  {NoStop}%
\bibitem [{\citenamefont {Saffman}\ \emph {et~al.}(2010)\citenamefont
  {Saffman}, \citenamefont {Walker},\ and\ \citenamefont
  {M\o{}lmer}}]{Saffman2010}%
  \BibitemOpen
  \bibfield  {author} {\bibinfo {author} {\bibfnamefont {M.}~\bibnamefont
  {Saffman}}, \bibinfo {author} {\bibfnamefont {T.~G.}\ \bibnamefont {Walker}},
  \ and\ \bibinfo {author} {\bibfnamefont {K.}~\bibnamefont {M\o{}lmer}},\
  }\href {\doibase 10.1103/RevModPhys.82.2313} {\bibfield  {journal} {\bibinfo
  {journal} {Rev. Mod. Phys.}\ }\textbf {\bibinfo {volume} {82}},\ \bibinfo
  {pages} {2313} (\bibinfo {year} {2010})}\BibitemShut {NoStop}%
\bibitem [{\citenamefont {Browaeys}\ and\ \citenamefont
  {Lahaye}(2020)}]{Browaeys2020}%
  \BibitemOpen
  \bibfield  {author} {\bibinfo {author} {\bibfnamefont {A.}~\bibnamefont
  {Browaeys}}\ and\ \bibinfo {author} {\bibfnamefont {T.}~\bibnamefont
  {Lahaye}},\ }\href {\doibase 10.1038/s41567-019-0733-z} {\bibfield  {journal}
  {\bibinfo  {journal} {Nature Physics}\ }\textbf {\bibinfo {volume} {16}},\
  \bibinfo {pages} {132} (\bibinfo {year} {2020})}\BibitemShut {NoStop}%
\bibitem [{\citenamefont {Barredo}\ \emph {et~al.}(2016)\citenamefont
  {Barredo}, \citenamefont {de~L{\'e}s{\'e}leuc}, \citenamefont {Lienhard},
  \citenamefont {Lahaye},\ and\ \citenamefont {Browaeys}}]{Barredo2016}%
  \BibitemOpen
  \bibfield  {author} {\bibinfo {author} {\bibfnamefont {D.}~\bibnamefont
  {Barredo}}, \bibinfo {author} {\bibfnamefont {S.}~\bibnamefont
  {de~L{\'e}s{\'e}leuc}}, \bibinfo {author} {\bibfnamefont {V.}~\bibnamefont
  {Lienhard}}, \bibinfo {author} {\bibfnamefont {T.}~\bibnamefont {Lahaye}}, \
  and\ \bibinfo {author} {\bibfnamefont {A.}~\bibnamefont {Browaeys}},\ }\href
  {\doibase 10.1126/science.aah3778} {\bibfield  {journal} {\bibinfo  {journal}
  {Science}\ } (\bibinfo {year} {2016}),\ 10.1126/science.aah3778}\BibitemShut
  {NoStop}%
\bibitem [{\citenamefont {Bernien}\ \emph {et~al.}(2017)\citenamefont
  {Bernien}, \citenamefont {Schwartz}, \citenamefont {Keesling}, \citenamefont
  {Levine}, \citenamefont {Omran}, \citenamefont {Pichler}, \citenamefont
  {Choi}, \citenamefont {Zibrov}, \citenamefont {Endres}, \citenamefont
  {Greiner}, \citenamefont {Vuleti{\'{c}}},\ and\ \citenamefont
  {Lukin}}]{Bernien2017}%
  \BibitemOpen
  \bibfield  {author} {\bibinfo {author} {\bibfnamefont {H.}~\bibnamefont
  {Bernien}}, \bibinfo {author} {\bibfnamefont {S.}~\bibnamefont {Schwartz}},
  \bibinfo {author} {\bibfnamefont {A.}~\bibnamefont {Keesling}}, \bibinfo
  {author} {\bibfnamefont {H.}~\bibnamefont {Levine}}, \bibinfo {author}
  {\bibfnamefont {A.}~\bibnamefont {Omran}}, \bibinfo {author} {\bibfnamefont
  {H.}~\bibnamefont {Pichler}}, \bibinfo {author} {\bibfnamefont
  {S.}~\bibnamefont {Choi}}, \bibinfo {author} {\bibfnamefont {A.~S.}\
  \bibnamefont {Zibrov}}, \bibinfo {author} {\bibfnamefont {M.}~\bibnamefont
  {Endres}}, \bibinfo {author} {\bibfnamefont {M.}~\bibnamefont {Greiner}},
  \bibinfo {author} {\bibfnamefont {V.}~\bibnamefont {Vuleti{\'{c}}}}, \ and\
  \bibinfo {author} {\bibfnamefont {M.~D.}\ \bibnamefont {Lukin}},\ }\href
  {\doibase 10.1038/nature24622} {\bibfield  {journal} {\bibinfo  {journal}
  {Nature}\ }\textbf {\bibinfo {volume} {551}},\ \bibinfo {pages} {579}
  (\bibinfo {year} {2017})}\BibitemShut {NoStop}%
\bibitem [{\citenamefont {Barredo}\ \emph {et~al.}(2018)\citenamefont
  {Barredo}, \citenamefont {Lienhard}, \citenamefont {de~L{\'e}s{\'e}leuc},
  \citenamefont {Lahaye},\ and\ \citenamefont {Browaeys}}]{Barredo2018}%
  \BibitemOpen
  \bibfield  {author} {\bibinfo {author} {\bibfnamefont {D.}~\bibnamefont
  {Barredo}}, \bibinfo {author} {\bibfnamefont {V.}~\bibnamefont {Lienhard}},
  \bibinfo {author} {\bibfnamefont {S.}~\bibnamefont {de~L{\'e}s{\'e}leuc}},
  \bibinfo {author} {\bibfnamefont {T.}~\bibnamefont {Lahaye}}, \ and\ \bibinfo
  {author} {\bibfnamefont {A.}~\bibnamefont {Browaeys}},\ }\href {\doibase
  10.1038/s41586-018-0450-2} {\bibfield  {journal} {\bibinfo  {journal}
  {Nature}\ }\textbf {\bibinfo {volume} {561}},\ \bibinfo {pages} {79}
  (\bibinfo {year} {2018})}\BibitemShut {NoStop}%
\bibitem [{\citenamefont {Schymik}\ \emph {et~al.}(2020)\citenamefont
  {Schymik}, \citenamefont {Lienhard}, \citenamefont {Barredo}, \citenamefont
  {Scholl}, \citenamefont {Williams}, \citenamefont {Browaeys},\ and\
  \citenamefont {Lahaye}}]{Schmymik2020}%
  \BibitemOpen
  \bibfield  {author} {\bibinfo {author} {\bibfnamefont {K.-N.}\ \bibnamefont
  {Schymik}}, \bibinfo {author} {\bibfnamefont {V.}~\bibnamefont {Lienhard}},
  \bibinfo {author} {\bibfnamefont {D.}~\bibnamefont {Barredo}}, \bibinfo
  {author} {\bibfnamefont {P.}~\bibnamefont {Scholl}}, \bibinfo {author}
  {\bibfnamefont {H.}~\bibnamefont {Williams}}, \bibinfo {author}
  {\bibfnamefont {A.}~\bibnamefont {Browaeys}}, \ and\ \bibinfo {author}
  {\bibfnamefont {T.}~\bibnamefont {Lahaye}},\ }\href {\doibase
  10.1103/PhysRevA.102.063107} {\bibfield  {journal} {\bibinfo  {journal}
  {Phys. Rev. A}\ }\textbf {\bibinfo {volume} {102}},\ \bibinfo {pages}
  {063107} (\bibinfo {year} {2020})}\BibitemShut {NoStop}%
\bibitem [{\citenamefont {Saffman}\ and\ \citenamefont
  {M\o{}lmer}(2008)}]{Saffman2008}%
  \BibitemOpen
  \bibfield  {author} {\bibinfo {author} {\bibfnamefont {M.}~\bibnamefont
  {Saffman}}\ and\ \bibinfo {author} {\bibfnamefont {K.}~\bibnamefont
  {M\o{}lmer}},\ }\href {\doibase 10.1103/PhysRevA.78.012336} {\bibfield
  {journal} {\bibinfo  {journal} {Phys. Rev. A}\ }\textbf {\bibinfo {volume}
  {78}},\ \bibinfo {pages} {012336} (\bibinfo {year} {2008})}\BibitemShut
  {NoStop}%
\bibitem [{\citenamefont {Mukherjee}\ \emph {et~al.}(2011)\citenamefont
  {Mukherjee}, \citenamefont {Millen}, \citenamefont {Nath}, \citenamefont
  {Jones},\ and\ \citenamefont {Pohl}}]{Mukherjee2011}%
  \BibitemOpen
  \bibfield  {author} {\bibinfo {author} {\bibfnamefont {R.}~\bibnamefont
  {Mukherjee}}, \bibinfo {author} {\bibfnamefont {J.}~\bibnamefont {Millen}},
  \bibinfo {author} {\bibfnamefont {R.}~\bibnamefont {Nath}}, \bibinfo {author}
  {\bibfnamefont {M.~P.~A.}\ \bibnamefont {Jones}}, \ and\ \bibinfo {author}
  {\bibfnamefont {T.}~\bibnamefont {Pohl}},\ }\href {\doibase
  10.1088/0953-4075/44/18/184010} {\bibfield  {journal} {\bibinfo  {journal}
  {J. Phys. B: At. Mol. Opt. Phys.}\ }\textbf {\bibinfo {volume} {44}},\
  \bibinfo {pages} {184010} (\bibinfo {year} {2011})}\BibitemShut {NoStop}%
\bibitem [{\citenamefont {Topcu}\ and\ \citenamefont
  {Derevianko}(2014)}]{Topcu2014}%
  \BibitemOpen
  \bibfield  {author} {\bibinfo {author} {\bibfnamefont {T.}~\bibnamefont
  {Topcu}}\ and\ \bibinfo {author} {\bibfnamefont {A.}~\bibnamefont
  {Derevianko}},\ }\href {\doibase 10.1103/PhysRevA.89.023411} {\bibfield
  {journal} {\bibinfo  {journal} {Phys. Rev. A}\ }\textbf {\bibinfo {volume}
  {89}},\ \bibinfo {pages} {023411} (\bibinfo {year} {2014})}\BibitemShut
  {NoStop}%
\bibitem [{\citenamefont {Robicheaux}\ \emph
  {et~al.}(2018{\natexlab{a}})\citenamefont {Robicheaux}, \citenamefont
  {Booth},\ and\ \citenamefont {Saffman}}]{RBS2018}%
  \BibitemOpen
  \bibfield  {author} {\bibinfo {author} {\bibfnamefont {F.}~\bibnamefont
  {Robicheaux}}, \bibinfo {author} {\bibfnamefont {D.~W.}\ \bibnamefont
  {Booth}}, \ and\ \bibinfo {author} {\bibfnamefont {M.}~\bibnamefont
  {Saffman}},\ }\href {\doibase 10.1103/PhysRevA.97.022508} {\bibfield
  {journal} {\bibinfo  {journal} {Phys. Rev. A}\ }\textbf {\bibinfo {volume}
  {97}},\ \bibinfo {pages} {022508} (\bibinfo {year}
  {2018}{\natexlab{a}})}\BibitemShut {NoStop}%
\bibitem [{\citenamefont {Wilson}\ \emph {et~al.}(2019)\citenamefont {Wilson},
  \citenamefont {Saskin}, \citenamefont {Meng}, \citenamefont {Ma},
  \citenamefont {Dilip}, \citenamefont {Burgers},\ and\ \citenamefont
  {Thompson}}]{Wilson2019}%
  \BibitemOpen
  \bibfield  {author} {\bibinfo {author} {\bibfnamefont {J.}~\bibnamefont
  {Wilson}}, \bibinfo {author} {\bibfnamefont {S.}~\bibnamefont {Saskin}},
  \bibinfo {author} {\bibfnamefont {Y.}~\bibnamefont {Meng}}, \bibinfo {author}
  {\bibfnamefont {S.}~\bibnamefont {Ma}}, \bibinfo {author} {\bibfnamefont
  {R.}~\bibnamefont {Dilip}}, \bibinfo {author} {\bibfnamefont
  {A.}~\bibnamefont {Burgers}}, \ and\ \bibinfo {author} {\bibfnamefont
  {J.}~\bibnamefont {Thompson}},\ }\href@noop {} {\enquote {\bibinfo {title}
  {Trapped arrays of alkaline earth rydberg atoms in optical tweezers},}\ }
  (\bibinfo {year} {2019}),\ \Eprint {http://arxiv.org/abs/1912.08754}
  {arXiv:1912.08754 [quant-ph]} \BibitemShut {NoStop}%
\bibitem [{\citenamefont {Millen}\ \emph {et~al.}(2010)\citenamefont {Millen},
  \citenamefont {Lochead},\ and\ \citenamefont {Jones}}]{Millen2010}%
  \BibitemOpen
  \bibfield  {author} {\bibinfo {author} {\bibfnamefont {J.}~\bibnamefont
  {Millen}}, \bibinfo {author} {\bibfnamefont {G.}~\bibnamefont {Lochead}}, \
  and\ \bibinfo {author} {\bibfnamefont {M.~P.~A.}\ \bibnamefont {Jones}},\
  }\href {\doibase 10.1103/PhysRevLett.105.213004} {\bibfield  {journal}
  {\bibinfo  {journal} {Phys. Rev. Lett.}\ }\textbf {\bibinfo {volume} {105}},\
  \bibinfo {pages} {213004} (\bibinfo {year} {2010})}\BibitemShut {NoStop}%
\bibitem [{\citenamefont {Bounds}\ \emph {et~al.}(2018)\citenamefont {Bounds},
  \citenamefont {Jackson}, \citenamefont {Hanley}, \citenamefont {Faoro},
  \citenamefont {Bridge}, \citenamefont {Huillery},\ and\ \citenamefont
  {Jones}}]{Bounds2018}%
  \BibitemOpen
  \bibfield  {author} {\bibinfo {author} {\bibfnamefont {A.~D.}\ \bibnamefont
  {Bounds}}, \bibinfo {author} {\bibfnamefont {N.~C.}\ \bibnamefont {Jackson}},
  \bibinfo {author} {\bibfnamefont {R.~K.}\ \bibnamefont {Hanley}}, \bibinfo
  {author} {\bibfnamefont {R.}~\bibnamefont {Faoro}}, \bibinfo {author}
  {\bibfnamefont {E.~M.}\ \bibnamefont {Bridge}}, \bibinfo {author}
  {\bibfnamefont {P.}~\bibnamefont {Huillery}}, \ and\ \bibinfo {author}
  {\bibfnamefont {M.~P.~A.}\ \bibnamefont {Jones}},\ }\href {\doibase
  10.1103/PhysRevLett.120.183401} {\bibfield  {journal} {\bibinfo  {journal}
  {Phys. Rev. Lett.}\ }\textbf {\bibinfo {volume} {120}},\ \bibinfo {pages}
  {183401} (\bibinfo {year} {2018})}\BibitemShut {NoStop}%
\bibitem [{\citenamefont {Camargo}\ \emph {et~al.}(2018)\citenamefont
  {Camargo}, \citenamefont {Schmidt}, \citenamefont {Whalen}, \citenamefont
  {Ding}, \citenamefont {Woehl}, \citenamefont {Yoshida}, \citenamefont
  {Burgd\"orfer}, \citenamefont {Dunning}, \citenamefont {Sadeghpour},
  \citenamefont {Demler},\ and\ \citenamefont {Killian}}]{Camargo2018}%
  \BibitemOpen
  \bibfield  {author} {\bibinfo {author} {\bibfnamefont {F.}~\bibnamefont
  {Camargo}}, \bibinfo {author} {\bibfnamefont {R.}~\bibnamefont {Schmidt}},
  \bibinfo {author} {\bibfnamefont {J.~D.}\ \bibnamefont {Whalen}}, \bibinfo
  {author} {\bibfnamefont {R.}~\bibnamefont {Ding}}, \bibinfo {author}
  {\bibfnamefont {G.}~\bibnamefont {Woehl}}, \bibinfo {author} {\bibfnamefont
  {S.}~\bibnamefont {Yoshida}}, \bibinfo {author} {\bibfnamefont
  {J.}~\bibnamefont {Burgd\"orfer}}, \bibinfo {author} {\bibfnamefont {F.~B.}\
  \bibnamefont {Dunning}}, \bibinfo {author} {\bibfnamefont {H.~R.}\
  \bibnamefont {Sadeghpour}}, \bibinfo {author} {\bibfnamefont
  {E.}~\bibnamefont {Demler}}, \ and\ \bibinfo {author} {\bibfnamefont {T.~C.}\
  \bibnamefont {Killian}},\ }\href {\doibase 10.1103/PhysRevLett.120.083401}
  {\bibfield  {journal} {\bibinfo  {journal} {Phys. Rev. Lett.}\ }\textbf
  {\bibinfo {volume} {120}},\ \bibinfo {pages} {083401} (\bibinfo {year}
  {2018})}\BibitemShut {NoStop}%
\bibitem [{\citenamefont {Couturier}\ \emph {et~al.}(2019)\citenamefont
  {Couturier}, \citenamefont {Nosske}, \citenamefont {Hu}, \citenamefont {Tan},
  \citenamefont {Qiao}, \citenamefont {Jiang}, \citenamefont {Chen},\ and\
  \citenamefont {Weidem\"uller}}]{Couturier2019}%
  \BibitemOpen
  \bibfield  {author} {\bibinfo {author} {\bibfnamefont {L.}~\bibnamefont
  {Couturier}}, \bibinfo {author} {\bibfnamefont {I.}~\bibnamefont {Nosske}},
  \bibinfo {author} {\bibfnamefont {F.}~\bibnamefont {Hu}}, \bibinfo {author}
  {\bibfnamefont {C.}~\bibnamefont {Tan}}, \bibinfo {author} {\bibfnamefont
  {C.}~\bibnamefont {Qiao}}, \bibinfo {author} {\bibfnamefont {Y.~H.}\
  \bibnamefont {Jiang}}, \bibinfo {author} {\bibfnamefont {P.}~\bibnamefont
  {Chen}}, \ and\ \bibinfo {author} {\bibfnamefont {M.}~\bibnamefont
  {Weidem\"uller}},\ }\href {\doibase 10.1103/PhysRevA.99.022503} {\bibfield
  {journal} {\bibinfo  {journal} {Phys. Rev. A}\ }\textbf {\bibinfo {volume}
  {99}},\ \bibinfo {pages} {022503} (\bibinfo {year} {2019})}\BibitemShut
  {NoStop}%
\bibitem [{\citenamefont {Lehec}\ \emph {et~al.}(2018)\citenamefont {Lehec},
  \citenamefont {Zuliani}, \citenamefont {Maineult}, \citenamefont
  {Luc-Koenig}, \citenamefont {Pillet}, \citenamefont {Cheinet}, \citenamefont
  {Niyaz},\ and\ \citenamefont {Gallagher}}]{Lehec2018}%
  \BibitemOpen
  \bibfield  {author} {\bibinfo {author} {\bibfnamefont {H.}~\bibnamefont
  {Lehec}}, \bibinfo {author} {\bibfnamefont {A.}~\bibnamefont {Zuliani}},
  \bibinfo {author} {\bibfnamefont {W.}~\bibnamefont {Maineult}}, \bibinfo
  {author} {\bibfnamefont {E.}~\bibnamefont {Luc-Koenig}}, \bibinfo {author}
  {\bibfnamefont {P.}~\bibnamefont {Pillet}}, \bibinfo {author} {\bibfnamefont
  {P.}~\bibnamefont {Cheinet}}, \bibinfo {author} {\bibfnamefont
  {F.}~\bibnamefont {Niyaz}}, \ and\ \bibinfo {author} {\bibfnamefont {T.~F.}\
  \bibnamefont {Gallagher}},\ }\href {\doibase 10.1103/PhysRevA.98.062506}
  {\bibfield  {journal} {\bibinfo  {journal} {Phys. Rev. A}\ }\textbf {\bibinfo
  {volume} {98}},\ \bibinfo {pages} {062506} (\bibinfo {year}
  {2018})}\BibitemShut {NoStop}%
\bibitem [{\citenamefont {Norcia}\ \emph {et~al.}(2018)\citenamefont {Norcia},
  \citenamefont {Young},\ and\ \citenamefont {Kaufman}}]{Norcia2018}%
  \BibitemOpen
  \bibfield  {author} {\bibinfo {author} {\bibfnamefont {M.~A.}\ \bibnamefont
  {Norcia}}, \bibinfo {author} {\bibfnamefont {A.~W.}\ \bibnamefont {Young}}, \
  and\ \bibinfo {author} {\bibfnamefont {A.~M.}\ \bibnamefont {Kaufman}},\
  }\href {\doibase 10.1103/PhysRevX.8.041054} {\bibfield  {journal} {\bibinfo
  {journal} {Phys. Rev. X}\ }\textbf {\bibinfo {volume} {8}},\ \bibinfo {pages}
  {041054} (\bibinfo {year} {2018})}\BibitemShut {NoStop}%
\bibitem [{\citenamefont {Cooper}\ \emph {et~al.}(2018)\citenamefont {Cooper},
  \citenamefont {Covey}, \citenamefont {Madjarov}, \citenamefont {Porsev},
  \citenamefont {Safronova},\ and\ \citenamefont {Endres}}]{Cooper2018}%
  \BibitemOpen
  \bibfield  {author} {\bibinfo {author} {\bibfnamefont {A.}~\bibnamefont
  {Cooper}}, \bibinfo {author} {\bibfnamefont {J.~P.}\ \bibnamefont {Covey}},
  \bibinfo {author} {\bibfnamefont {I.~S.}\ \bibnamefont {Madjarov}}, \bibinfo
  {author} {\bibfnamefont {S.~G.}\ \bibnamefont {Porsev}}, \bibinfo {author}
  {\bibfnamefont {M.~S.}\ \bibnamefont {Safronova}}, \ and\ \bibinfo {author}
  {\bibfnamefont {M.}~\bibnamefont {Endres}},\ }\href {\doibase
  10.1103/PhysRevX.8.041055} {\bibfield  {journal} {\bibinfo  {journal} {Phys.
  Rev. X}\ }\textbf {\bibinfo {volume} {8}},\ \bibinfo {pages} {041055}
  (\bibinfo {year} {2018})}\BibitemShut {NoStop}%
\bibitem [{\citenamefont {Saskin}\ \emph {et~al.}(2019)\citenamefont {Saskin},
  \citenamefont {Wilson}, \citenamefont {Grinkemeyer},\ and\ \citenamefont
  {Thompson}}]{Saskin2019}%
  \BibitemOpen
  \bibfield  {author} {\bibinfo {author} {\bibfnamefont {S.}~\bibnamefont
  {Saskin}}, \bibinfo {author} {\bibfnamefont {J.~T.}\ \bibnamefont {Wilson}},
  \bibinfo {author} {\bibfnamefont {B.}~\bibnamefont {Grinkemeyer}}, \ and\
  \bibinfo {author} {\bibfnamefont {J.~D.}\ \bibnamefont {Thompson}},\ }\href
  {\doibase 10.1103/PhysRevLett.122.143002} {\bibfield  {journal} {\bibinfo
  {journal} {Phys. Rev. Lett.}\ }\textbf {\bibinfo {volume} {122}},\ \bibinfo
  {pages} {143002} (\bibinfo {year} {2019})}\BibitemShut {NoStop}%
\bibitem [{\citenamefont {Madjarov}\ \emph {et~al.}(2020)\citenamefont
  {Madjarov}, \citenamefont {Covey}, \citenamefont {Shaw}, \citenamefont
  {Choi}, \citenamefont {Kale}, \citenamefont {Cooper}, \citenamefont
  {Pichler}, \citenamefont {Schkolnik}, \citenamefont {Williams},\ and\
  \citenamefont {Endres}}]{Madjarov2020}%
  \BibitemOpen
  \bibfield  {author} {\bibinfo {author} {\bibfnamefont {I.~S.}\ \bibnamefont
  {Madjarov}}, \bibinfo {author} {\bibfnamefont {J.~P.}\ \bibnamefont {Covey}},
  \bibinfo {author} {\bibfnamefont {A.~L.}\ \bibnamefont {Shaw}}, \bibinfo
  {author} {\bibfnamefont {J.}~\bibnamefont {Choi}}, \bibinfo {author}
  {\bibfnamefont {A.}~\bibnamefont {Kale}}, \bibinfo {author} {\bibfnamefont
  {A.}~\bibnamefont {Cooper}}, \bibinfo {author} {\bibfnamefont
  {H.}~\bibnamefont {Pichler}}, \bibinfo {author} {\bibfnamefont
  {V.}~\bibnamefont {Schkolnik}}, \bibinfo {author} {\bibfnamefont {J.~R.}\
  \bibnamefont {Williams}}, \ and\ \bibinfo {author} {\bibfnamefont
  {M.}~\bibnamefont {Endres}},\ }\href {\doibase 10.1038/s41567-020-0903-z}
  {\bibfield  {journal} {\bibinfo  {journal} {Nature Physics}\ }\textbf
  {\bibinfo {volume} {16}},\ \bibinfo {pages} {857} (\bibinfo {year}
  {2020})}\BibitemShut {NoStop}%
\bibitem [{\citenamefont {McClelland}\ and\ \citenamefont
  {Hanssen}(2006)}]{McClelland2006}%
  \BibitemOpen
  \bibfield  {author} {\bibinfo {author} {\bibfnamefont {J.~J.}\ \bibnamefont
  {McClelland}}\ and\ \bibinfo {author} {\bibfnamefont {J.~L.}\ \bibnamefont
  {Hanssen}},\ }\href {\doibase 10.1103/PhysRevLett.96.143005} {\bibfield
  {journal} {\bibinfo  {journal} {Phys. Rev. Lett.}\ }\textbf {\bibinfo
  {volume} {96}},\ \bibinfo {pages} {143005} (\bibinfo {year}
  {2006})}\BibitemShut {NoStop}%
\bibitem [{\citenamefont {Lev}\ \emph {et~al.}(2010)\citenamefont {Lev},
  \citenamefont {Lu},\ and\ \citenamefont {Youn}}]{Lev2010}%
  \BibitemOpen
  \bibfield  {author} {\bibinfo {author} {\bibfnamefont {B.}~\bibnamefont
  {Lev}}, \bibinfo {author} {\bibfnamefont {M.}~\bibnamefont {Lu}}, \ and\
  \bibinfo {author} {\bibfnamefont {S.~H.}\ \bibnamefont {Youn}},\ }in\ \href
  {\doibase 10.1364/FIO.2010.STuD4} {\emph {\bibinfo {booktitle} {Frontiers in
  Optics 2010/Laser Science XXVI}}}\ (\bibinfo  {publisher} {Optical Society of
  America},\ \bibinfo {year} {2010})\ p.\ \bibinfo {pages} {STuD4}\BibitemShut
  {NoStop}%
\bibitem [{\citenamefont {Frisch}\ \emph {et~al.}(2012)\citenamefont {Frisch},
  \citenamefont {Aikawa}, \citenamefont {Mark}, \citenamefont {Rietzler},
  \citenamefont {Schindler}, \citenamefont {Zupani\ifmmode~\check{c}\else
  \v{c}\fi{}}, \citenamefont {Grimm},\ and\ \citenamefont
  {Ferlaino}}]{Frisch2012}%
  \BibitemOpen
  \bibfield  {author} {\bibinfo {author} {\bibfnamefont {A.}~\bibnamefont
  {Frisch}}, \bibinfo {author} {\bibfnamefont {K.}~\bibnamefont {Aikawa}},
  \bibinfo {author} {\bibfnamefont {M.}~\bibnamefont {Mark}}, \bibinfo {author}
  {\bibfnamefont {A.}~\bibnamefont {Rietzler}}, \bibinfo {author}
  {\bibfnamefont {J.}~\bibnamefont {Schindler}}, \bibinfo {author}
  {\bibfnamefont {E.}~\bibnamefont {Zupani\ifmmode~\check{c}\else \v{c}\fi{}}},
  \bibinfo {author} {\bibfnamefont {R.}~\bibnamefont {Grimm}}, \ and\ \bibinfo
  {author} {\bibfnamefont {F.}~\bibnamefont {Ferlaino}},\ }\href {\doibase
  10.1103/PhysRevA.85.051401} {\bibfield  {journal} {\bibinfo  {journal} {Phys.
  Rev. A}\ }\textbf {\bibinfo {volume} {85}},\ \bibinfo {pages} {051401(R)}
  (\bibinfo {year} {2012})}\BibitemShut {NoStop}%
\bibitem [{\citenamefont {Miao}\ \emph {et~al.}(2014)\citenamefont {Miao},
  \citenamefont {Hostetter}, \citenamefont {Stratis},\ and\ \citenamefont
  {Saffman}}]{Miao2014}%
  \BibitemOpen
  \bibfield  {author} {\bibinfo {author} {\bibfnamefont {J.}~\bibnamefont
  {Miao}}, \bibinfo {author} {\bibfnamefont {J.}~\bibnamefont {Hostetter}},
  \bibinfo {author} {\bibfnamefont {G.}~\bibnamefont {Stratis}}, \ and\
  \bibinfo {author} {\bibfnamefont {M.}~\bibnamefont {Saffman}},\ }\href
  {\doibase 10.1103/PhysRevA.89.041401} {\bibfield  {journal} {\bibinfo
  {journal} {Phys. Rev. A}\ }\textbf {\bibinfo {volume} {89}},\ \bibinfo
  {pages} {041401(R)} (\bibinfo {year} {2014})}\BibitemShut {NoStop}%
\bibitem [{\citenamefont {Cojocaru}\ \emph {et~al.}(2017)\citenamefont
  {Cojocaru}, \citenamefont {Pyatchenkov}, \citenamefont {Snigirev},
  \citenamefont {Luchnikov}, \citenamefont {Kalganova}, \citenamefont
  {Vishnyakova}, \citenamefont {Kublikova}, \citenamefont {Bushmakin},
  \citenamefont {Davletov}, \citenamefont {Tsyganok}, \citenamefont {Belyaeva},
  \citenamefont {Khoroshilov}, \citenamefont {Sorokin}, \citenamefont
  {Sukachev},\ and\ \citenamefont {Akimov}}]{Cojocaru2017}%
  \BibitemOpen
  \bibfield  {author} {\bibinfo {author} {\bibfnamefont {I.~S.}\ \bibnamefont
  {Cojocaru}}, \bibinfo {author} {\bibfnamefont {S.~V.}\ \bibnamefont
  {Pyatchenkov}}, \bibinfo {author} {\bibfnamefont {S.~A.}\ \bibnamefont
  {Snigirev}}, \bibinfo {author} {\bibfnamefont {I.~A.}\ \bibnamefont
  {Luchnikov}}, \bibinfo {author} {\bibfnamefont {E.~S.}\ \bibnamefont
  {Kalganova}}, \bibinfo {author} {\bibfnamefont {G.~A.}\ \bibnamefont
  {Vishnyakova}}, \bibinfo {author} {\bibfnamefont {D.~N.}\ \bibnamefont
  {Kublikova}}, \bibinfo {author} {\bibfnamefont {V.~S.}\ \bibnamefont
  {Bushmakin}}, \bibinfo {author} {\bibfnamefont {E.~T.}\ \bibnamefont
  {Davletov}}, \bibinfo {author} {\bibfnamefont {V.~V.}\ \bibnamefont
  {Tsyganok}}, \bibinfo {author} {\bibfnamefont {O.~V.}\ \bibnamefont
  {Belyaeva}}, \bibinfo {author} {\bibfnamefont {A.}~\bibnamefont
  {Khoroshilov}}, \bibinfo {author} {\bibfnamefont {V.~N.}\ \bibnamefont
  {Sorokin}}, \bibinfo {author} {\bibfnamefont {D.~D.}\ \bibnamefont
  {Sukachev}}, \ and\ \bibinfo {author} {\bibfnamefont {A.~V.}\ \bibnamefont
  {Akimov}},\ }\href {\doibase 10.1103/PhysRevA.95.012706} {\bibfield
  {journal} {\bibinfo  {journal} {Phys. Rev. A}\ }\textbf {\bibinfo {volume}
  {95}},\ \bibinfo {pages} {012706} (\bibinfo {year} {2017})}\BibitemShut
  {NoStop}%
\bibitem [{\citenamefont {Robicheaux}\ \emph
  {et~al.}(2018{\natexlab{b}})\citenamefont {Robicheaux}, \citenamefont
  {Booth},\ and\ \citenamefont {Saffman}}]{Robicheaux2018}%
  \BibitemOpen
  \bibfield  {author} {\bibinfo {author} {\bibfnamefont {F.}~\bibnamefont
  {Robicheaux}}, \bibinfo {author} {\bibfnamefont {D.~W.}\ \bibnamefont
  {Booth}}, \ and\ \bibinfo {author} {\bibfnamefont {M.}~\bibnamefont
  {Saffman}},\ }\href {\doibase 10.1103/PhysRevA.97.022508} {\bibfield
  {journal} {\bibinfo  {journal} {Phys. Rev. A}\ }\textbf {\bibinfo {volume}
  {97}},\ \bibinfo {pages} {022508} (\bibinfo {year}
  {2018}{\natexlab{b}})}\BibitemShut {NoStop}%
\bibitem [{\citenamefont {Hostetter}\ \emph {et~al.}(2015)\citenamefont
  {Hostetter}, \citenamefont {Pritchard}, \citenamefont {Lawler},\ and\
  \citenamefont {Saffman}}]{Hostetter2015}%
  \BibitemOpen
  \bibfield  {author} {\bibinfo {author} {\bibfnamefont {J.}~\bibnamefont
  {Hostetter}}, \bibinfo {author} {\bibfnamefont {J.~D.}\ \bibnamefont
  {Pritchard}}, \bibinfo {author} {\bibfnamefont {J.~E.}\ \bibnamefont
  {Lawler}}, \ and\ \bibinfo {author} {\bibfnamefont {M.}~\bibnamefont
  {Saffman}},\ }\href {\doibase 10.1103/PhysRevA.91.012507} {\bibfield
  {journal} {\bibinfo  {journal} {Phys. Rev. A}\ }\textbf {\bibinfo {volume}
  {91}},\ \bibinfo {pages} {012507} (\bibinfo {year} {2015})}\BibitemShut
  {NoStop}%
\bibitem [{\citenamefont {Hai-jun}\ \emph {et~al.}(1992)\citenamefont
  {Hai-jun}, \citenamefont {Xiang-yuan}, \citenamefont {Wen}, \citenamefont
  {Liang-quan},\ and\ \citenamefont {Die-yan}}]{Haijun1992}%
  \BibitemOpen
  \bibfield  {author} {\bibinfo {author} {\bibfnamefont {Z.}~\bibnamefont
  {Hai-jun}}, \bibinfo {author} {\bibfnamefont {X.}~\bibnamefont {Xiang-yuan}},
  \bibinfo {author} {\bibfnamefont {H.}~\bibnamefont {Wen}}, \bibinfo {author}
  {\bibfnamefont {L.}~\bibnamefont {Liang-quan}}, \ and\ \bibinfo {author}
  {\bibfnamefont {C.}~\bibnamefont {Die-yan}},\ }\href {\doibase
  10.1088/1004-423X/1/1/003} {\bibfield  {journal} {\bibinfo  {journal} {Acta
  Phys. Sin. (Overseas Edn)}\ }\textbf {\bibinfo {volume} {1}},\ \bibinfo
  {pages} {19} (\bibinfo {year} {1992})}\BibitemShut {NoStop}%
\bibitem [{\citenamefont {Studer}\ \emph {et~al.}(2016)\citenamefont {Studer},
  \citenamefont {Dyrauf}, \citenamefont {Naubereit}, \citenamefont {Heinke},\
  and\ \citenamefont {Wendt}}]{Studer2016}%
  \BibitemOpen
  \bibfield  {author} {\bibinfo {author} {\bibfnamefont {D.}~\bibnamefont
  {Studer}}, \bibinfo {author} {\bibfnamefont {P.}~\bibnamefont {Dyrauf}},
  \bibinfo {author} {\bibfnamefont {P.}~\bibnamefont {Naubereit}}, \bibinfo
  {author} {\bibfnamefont {R.}~\bibnamefont {Heinke}}, \ and\ \bibinfo {author}
  {\bibfnamefont {K.}~\bibnamefont {Wendt}},\ }\href {\doibase
  10.1007/s10751-016-1384-4} {\bibfield  {journal} {\bibinfo  {journal}
  {Hyperfine Interactions}\ }\textbf {\bibinfo {volume} {238}},\ \bibinfo
  {pages} {8} (\bibinfo {year} {2016})}\BibitemShut {NoStop}%
\bibitem [{\citenamefont {Studer}\ \emph {et~al.}(2019)\citenamefont {Studer},
  \citenamefont {Heinitz}, \citenamefont {Heinke}, \citenamefont {Naubereit},
  \citenamefont {Dressler}, \citenamefont {Guerrero}, \citenamefont {K\"oster},
  \citenamefont {Schumann},\ and\ \citenamefont {Wendt}}]{Studer2019}%
  \BibitemOpen
  \bibfield  {author} {\bibinfo {author} {\bibfnamefont {D.}~\bibnamefont
  {Studer}}, \bibinfo {author} {\bibfnamefont {S.}~\bibnamefont {Heinitz}},
  \bibinfo {author} {\bibfnamefont {R.}~\bibnamefont {Heinke}}, \bibinfo
  {author} {\bibfnamefont {P.}~\bibnamefont {Naubereit}}, \bibinfo {author}
  {\bibfnamefont {R.}~\bibnamefont {Dressler}}, \bibinfo {author}
  {\bibfnamefont {C.}~\bibnamefont {Guerrero}}, \bibinfo {author}
  {\bibfnamefont {U.}~\bibnamefont {K\"oster}}, \bibinfo {author}
  {\bibfnamefont {D.}~\bibnamefont {Schumann}}, \ and\ \bibinfo {author}
  {\bibfnamefont {K.}~\bibnamefont {Wendt}},\ }\href {\doibase
  10.1103/PhysRevA.99.062513} {\bibfield  {journal} {\bibinfo  {journal} {Phys.
  Rev. A}\ }\textbf {\bibinfo {volume} {99}},\ \bibinfo {pages} {062513}
  (\bibinfo {year} {2019})}\BibitemShut {NoStop}%
\bibitem [{\citenamefont {Studer}(2015)}]{Studer2015}%
  \BibitemOpen
  \bibfield  {author} {\bibinfo {author} {\bibfnamefont {D.}~\bibnamefont
  {Studer}},\ }\href@noop {} {\enquote {\bibinfo {title}
  {{Resonanzionisationsspektroskopie hoch\-liegender Zustände in Dysprosium
  und Erbium zur Entwicklung effizienter Anregungsschemata und Bestimmung des
  ersten Ionisationspotentials}},}\ } (\bibinfo {year} {2015}),\ \bibinfo
  {note} {master's thesis, private communication}\BibitemShut {NoStop}%
\bibitem [{\citenamefont {Boller}\ \emph {et~al.}(1991)\citenamefont {Boller},
  \citenamefont {Imamo\ifmmode~\breve{g}\else \u{g}\fi{}lu},\ and\
  \citenamefont {Harris}}]{Boller1991}%
  \BibitemOpen
  \bibfield  {author} {\bibinfo {author} {\bibfnamefont {K.-J.}\ \bibnamefont
  {Boller}}, \bibinfo {author} {\bibfnamefont {A.}~\bibnamefont
  {Imamo\ifmmode~\breve{g}\else \u{g}\fi{}lu}}, \ and\ \bibinfo {author}
  {\bibfnamefont {S.~E.}\ \bibnamefont {Harris}},\ }\href {\doibase
  10.1103/PhysRevLett.66.2593} {\bibfield  {journal} {\bibinfo  {journal}
  {Phys. Rev. Lett.}\ }\textbf {\bibinfo {volume} {66}},\ \bibinfo {pages}
  {2593} (\bibinfo {year} {1991})}\BibitemShut {NoStop}%
\bibitem [{\citenamefont {Mohapatra}\ \emph {et~al.}(2007)\citenamefont
  {Mohapatra}, \citenamefont {Jackson},\ and\ \citenamefont
  {Adams}}]{Mohapatra2007}%
  \BibitemOpen
  \bibfield  {author} {\bibinfo {author} {\bibfnamefont {A.~K.}\ \bibnamefont
  {Mohapatra}}, \bibinfo {author} {\bibfnamefont {T.~R.}\ \bibnamefont
  {Jackson}}, \ and\ \bibinfo {author} {\bibfnamefont {C.~S.}\ \bibnamefont
  {Adams}},\ }\href {\doibase 10.1103/PhysRevLett.98.113003} {\bibfield
  {journal} {\bibinfo  {journal} {Phys. Rev. Lett.}\ }\textbf {\bibinfo
  {volume} {98}},\ \bibinfo {pages} {113003} (\bibinfo {year}
  {2007})}\BibitemShut {NoStop}%
\bibitem [{\citenamefont {Mauger}\ \emph {et~al.}(2007)\citenamefont {Mauger},
  \citenamefont {Millen},\ and\ \citenamefont {Jones}}]{Mauger2007}%
  \BibitemOpen
  \bibfield  {author} {\bibinfo {author} {\bibfnamefont {S.}~\bibnamefont
  {Mauger}}, \bibinfo {author} {\bibfnamefont {J.}~\bibnamefont {Millen}}, \
  and\ \bibinfo {author} {\bibfnamefont {M.~P.~A.}\ \bibnamefont {Jones}},\
  }\href {\doibase 10.1088/0953-4075/40/22/f03} {\bibfield  {journal} {\bibinfo
   {journal} {Journal of Physics B: Atomic, Molecular and Optical Physics}\
  }\textbf {\bibinfo {volume} {40}},\ \bibinfo {pages} {F319} (\bibinfo {year}
  {2007})}\BibitemShut {NoStop}%
\bibitem [{\citenamefont {Naber}\ \emph {et~al.}(2017)\citenamefont {Naber},
  \citenamefont {Tauschinsky}, \citenamefont {van Linden van~den Heuvell},\
  and\ \citenamefont {Spreeuw}}]{Naber2017}%
  \BibitemOpen
  \bibfield  {author} {\bibinfo {author} {\bibfnamefont {J.~B.}\ \bibnamefont
  {Naber}}, \bibinfo {author} {\bibfnamefont {A.}~\bibnamefont {Tauschinsky}},
  \bibinfo {author} {\bibfnamefont {H.~B.}\ \bibnamefont {van Linden van~den
  Heuvell}}, \ and\ \bibinfo {author} {\bibfnamefont {R.~J.~C.}\ \bibnamefont
  {Spreeuw}},\ }\href {\doibase 10.21468/SciPostPhys.2.2.015} {\bibfield
  {journal} {\bibinfo  {journal} {SciPost Phys.}\ }\textbf {\bibinfo {volume}
  {2}},\ \bibinfo {pages} {015} (\bibinfo {year} {2017})}\BibitemShut {NoStop}%
\bibitem [{\citenamefont {Lu}\ and\ \citenamefont {Fano}(1970)}]{LuFano1970}%
  \BibitemOpen
  \bibfield  {author} {\bibinfo {author} {\bibfnamefont {K.~T.}\ \bibnamefont
  {Lu}}\ and\ \bibinfo {author} {\bibfnamefont {U.}~\bibnamefont {Fano}},\
  }\href {\doibase 10.1103/PhysRevA.2.81} {\bibfield  {journal} {\bibinfo
  {journal} {Phys. Rev. A}\ }\textbf {\bibinfo {volume} {2}},\ \bibinfo {pages}
  {81} (\bibinfo {year} {1970})}\BibitemShut {NoStop}%
\bibitem [{\citenamefont {Vaillant}\ \emph {et~al.}(2014)\citenamefont
  {Vaillant}, \citenamefont {Jones},\ and\ \citenamefont
  {Potvliege}}]{Vaillant2014}%
  \BibitemOpen
  \bibfield  {author} {\bibinfo {author} {\bibfnamefont {C.~L.}\ \bibnamefont
  {Vaillant}}, \bibinfo {author} {\bibfnamefont {M.~P.~A.}\ \bibnamefont
  {Jones}}, \ and\ \bibinfo {author} {\bibfnamefont {R.~M.}\ \bibnamefont
  {Potvliege}},\ }\href {\doibase 10.1088/0953-4075/47/15/155001} {\bibfield
  {journal} {\bibinfo  {journal} {Journal of Physics B: Atomic, Molecular and
  Optical Physics}\ }\textbf {\bibinfo {volume} {47}},\ \bibinfo {pages}
  {155001} (\bibinfo {year} {2014})}\BibitemShut {NoStop}%
\bibitem [{\citenamefont {Ilzh\"ofer}\ \emph {et~al.}(2018)\citenamefont
  {Ilzh\"ofer}, \citenamefont {Durastante}, \citenamefont {Patscheider},
  \citenamefont {Trautmann}, \citenamefont {Mark},\ and\ \citenamefont
  {Ferlaino}}]{Ilzhofer2018}%
  \BibitemOpen
  \bibfield  {author} {\bibinfo {author} {\bibfnamefont {P.}~\bibnamefont
  {Ilzh\"ofer}}, \bibinfo {author} {\bibfnamefont {G.}~\bibnamefont
  {Durastante}}, \bibinfo {author} {\bibfnamefont {A.}~\bibnamefont
  {Patscheider}}, \bibinfo {author} {\bibfnamefont {A.}~\bibnamefont
  {Trautmann}}, \bibinfo {author} {\bibfnamefont {M.~J.}\ \bibnamefont {Mark}},
  \ and\ \bibinfo {author} {\bibfnamefont {F.}~\bibnamefont {Ferlaino}},\
  }\href {\doibase 10.1103/PhysRevA.97.023633} {\bibfield  {journal} {\bibinfo
  {journal} {Phys. Rev. A}\ }\textbf {\bibinfo {volume} {97}},\ \bibinfo
  {pages} {023633} (\bibinfo {year} {2018})}\BibitemShut {NoStop}%
\bibitem [{\citenamefont {Autler}\ and\ \citenamefont
  {Townes}(1955)}]{Autler1955}%
  \BibitemOpen
  \bibfield  {author} {\bibinfo {author} {\bibfnamefont {S.~H.}\ \bibnamefont
  {Autler}}\ and\ \bibinfo {author} {\bibfnamefont {C.~H.}\ \bibnamefont
  {Townes}},\ }\href {\doibase 10.1103/PhysRev.100.703} {\bibfield  {journal}
  {\bibinfo  {journal} {Phys. Rev.}\ }\textbf {\bibinfo {volume} {100}},\
  \bibinfo {pages} {703} (\bibinfo {year} {1955})}\BibitemShut {NoStop}%
\bibitem [{\citenamefont {Worden}\ \emph {et~al.}(1978)\citenamefont {Worden},
  \citenamefont {Solarz}, \citenamefont {Paisner},\ and\ \citenamefont
  {Conway}}]{Worden1978}%
  \BibitemOpen
  \bibfield  {author} {\bibinfo {author} {\bibfnamefont {E.~F.}\ \bibnamefont
  {Worden}}, \bibinfo {author} {\bibfnamefont {R.~W.}\ \bibnamefont {Solarz}},
  \bibinfo {author} {\bibfnamefont {J.~A.}\ \bibnamefont {Paisner}}, \ and\
  \bibinfo {author} {\bibfnamefont {J.~G.}\ \bibnamefont {Conway}},\ }\href
  {\doibase 10.1364/JOSA.68.000052} {\bibfield  {journal} {\bibinfo  {journal}
  {J. Opt. Soc. Am.}\ }\textbf {\bibinfo {volume} {68}},\ \bibinfo {pages} {52}
  (\bibinfo {year} {1978})}\BibitemShut {NoStop}%
\bibitem [{\citenamefont {Martin}\ \emph {et~al.}(1978)\citenamefont {Martin},
  \citenamefont {Zalubas},\ and\ \citenamefont {Hagan}}]{Martin1978}%
  \BibitemOpen
  \bibfield  {author} {\bibinfo {author} {\bibfnamefont {W.~C.}\ \bibnamefont
  {Martin}}, \bibinfo {author} {\bibfnamefont {R.}~\bibnamefont {Zalubas}}, \
  and\ \bibinfo {author} {\bibfnamefont {L.}~\bibnamefont {Hagan}},\ }\href
  {\doibase DOI:10.6028/NBS.NSRDS.60} {\bibfield  {journal} {\bibinfo
  {journal} {Nat. Stand. Ref. Data Ser. NSRDS-NBS 60}\ }\textbf {\bibinfo
  {volume} {60}} (\bibinfo {year} {1978}),\
  DOI:10.6028/NBS.NSRDS.60}\BibitemShut {NoStop}%
\bibitem [{\citenamefont {Wyart}\ and\ \citenamefont
  {Lawler}(2009)}]{Wyart2009}%
  \BibitemOpen
  \bibfield  {author} {\bibinfo {author} {\bibfnamefont {J.-F.}\ \bibnamefont
  {Wyart}}\ and\ \bibinfo {author} {\bibfnamefont {J.~E.}\ \bibnamefont
  {Lawler}},\ }\href {\doibase 10.1088/0031-8949/79/04/045301} {\bibfield
  {journal} {\bibinfo  {journal} {Physica Scripta}\ }\textbf {\bibinfo {volume}
  {79}},\ \bibinfo {pages} {045301} (\bibinfo {year} {2009})}\BibitemShut
  {NoStop}%
\bibitem [{Note1()}]{Note1}%
  \BibitemOpen
  \bibinfo {note} {For $n_\protect \text {eff} = 23$ we observe two $J=8$
  features very close together, which is not yet understood.}\BibitemShut
  {Stop}%
\bibitem [{\citenamefont {Frisch}\ \emph {et~al.}(2013)\citenamefont {Frisch},
  \citenamefont {Aikawa}, \citenamefont {Mark}, \citenamefont {Ferlaino},
  \citenamefont {Berseneva},\ and\ \citenamefont {Kotochigova}}]{Frisch2013}%
  \BibitemOpen
  \bibfield  {author} {\bibinfo {author} {\bibfnamefont {A.}~\bibnamefont
  {Frisch}}, \bibinfo {author} {\bibfnamefont {K.}~\bibnamefont {Aikawa}},
  \bibinfo {author} {\bibfnamefont {M.}~\bibnamefont {Mark}}, \bibinfo {author}
  {\bibfnamefont {F.}~\bibnamefont {Ferlaino}}, \bibinfo {author}
  {\bibfnamefont {E.}~\bibnamefont {Berseneva}}, \ and\ \bibinfo {author}
  {\bibfnamefont {S.}~\bibnamefont {Kotochigova}},\ }\href {\doibase
  10.1103/PhysRevA.88.032508} {\bibfield  {journal} {\bibinfo  {journal} {Phys.
  Rev. A}\ }\textbf {\bibinfo {volume} {88}},\ \bibinfo {pages} {032508}
  (\bibinfo {year} {2013})}\BibitemShut {NoStop}%
\bibitem [{\citenamefont {Aymar}\ \emph {et~al.}(1996)\citenamefont {Aymar},
  \citenamefont {Greene},\ and\ \citenamefont {Luc-Koenig}}]{AGL1996}%
  \BibitemOpen
  \bibfield  {author} {\bibinfo {author} {\bibfnamefont {M.}~\bibnamefont
  {Aymar}}, \bibinfo {author} {\bibfnamefont {C.~H.}\ \bibnamefont {Greene}}, \
  and\ \bibinfo {author} {\bibfnamefont {E.}~\bibnamefont {Luc-Koenig}},\
  }\href@noop {} {\bibfield  {journal} {\bibinfo  {journal} {\rmp}\ }\textbf
  {\bibinfo {volume} {68}},\ \bibinfo {pages} {1015} (\bibinfo {year}
  {1996})}\BibitemShut {NoStop}%
\bibitem [{\citenamefont {de~Laeter}\ \emph {et~al.}(2003)\citenamefont
  {de~Laeter}, \citenamefont {Böhlke}, \citenamefont {Bièvre}, \citenamefont
  {Hidaka}, \citenamefont {Peiser}, \citenamefont {Rosman},\ and\ \citenamefont
  {Taylor}}]{ErMass2000}%
  \BibitemOpen
  \bibfield  {author} {\bibinfo {author} {\bibfnamefont {J.~R.}\ \bibnamefont
  {de~Laeter}}, \bibinfo {author} {\bibfnamefont {J.~K.}\ \bibnamefont
  {Böhlke}}, \bibinfo {author} {\bibfnamefont {P.~D.}\ \bibnamefont
  {Bièvre}}, \bibinfo {author} {\bibfnamefont {H.}~\bibnamefont {Hidaka}},
  \bibinfo {author} {\bibfnamefont {H.~S.}\ \bibnamefont {Peiser}}, \bibinfo
  {author} {\bibfnamefont {K.~J.~R.}\ \bibnamefont {Rosman}}, \ and\ \bibinfo
  {author} {\bibfnamefont {P.~D.~P.}\ \bibnamefont {Taylor}},\ }\href {\doibase
  doi:10.1351/pac200375060683} {\bibfield  {journal} {\bibinfo  {journal} {Pure
  and Applied Chemistry}\ }\textbf {\bibinfo {volume} {75}},\ \bibinfo {pages}
  {683} (\bibinfo {year} {2003})}\BibitemShut {NoStop}%
\bibitem [{\citenamefont {Edmonds}(1974)}]{ARE1974}%
  \BibitemOpen
  \bibfield  {author} {\bibinfo {author} {\bibfnamefont {A.}~\bibnamefont
  {Edmonds}},\ }\href@noop {} {\emph {\bibinfo {title} {Angular Momentum in
  Quantum Mechanics, 2nd Edition}}}\ (\bibinfo  {publisher} {Princeton
  University Press, Princeton, New Jersey},\ \bibinfo {year}
  {1974})\BibitemShut {NoStop}%
\bibitem [{\citenamefont {Kramida}\ \emph {et~al.}(2020)\citenamefont
  {Kramida}, \citenamefont {{Yu.~Ralchenko}}, \citenamefont {Reader},\ and\
  \citenamefont {{and NIST ASD Team}}}]{NIST_ASD}%
  \BibitemOpen
  \bibfield  {author} {\bibinfo {author} {\bibfnamefont {A.}~\bibnamefont
  {Kramida}}, \bibinfo {author} {\bibnamefont {{Yu.~Ralchenko}}}, \bibinfo
  {author} {\bibfnamefont {J.}~\bibnamefont {Reader}}, \ and\ \bibinfo {author}
  {\bibnamefont {{and NIST ASD Team}}},\ }\href {\doibase 10.18434/T4W30F}
  {}\bibinfo {howpublished} {{NIST Atomic Spectra Database (ver. 5.8),
  [Online]. Available: {\tt{https://physics.nist.gov/asd}} [2021, February 17].
  National Institute of Standards and Technology, Gaithersburg, MD.}} (\bibinfo
  {year} {2020})\BibitemShut {NoStop}%
\bibitem [{\citenamefont {McNally}\ and\ \citenamefont
  {Sluis}(1959)}]{McNally1959}%
  \BibitemOpen
  \bibfield  {author} {\bibinfo {author} {\bibfnamefont {J.~R.}\ \bibnamefont
  {McNally}}\ and\ \bibinfo {author} {\bibfnamefont {K.~L.~V.}\ \bibnamefont
  {Sluis}},\ }\href {\doibase 10.1364/JOSA.49.000200} {\bibfield  {journal}
  {\bibinfo  {journal} {J. Opt. Soc. Am.}\ }\textbf {\bibinfo {volume} {49}},\
  \bibinfo {pages} {200} (\bibinfo {year} {1959})}\BibitemShut {NoStop}%
\bibitem [{\citenamefont {Becher}\ \emph {et~al.}(2018)\citenamefont {Becher},
  \citenamefont {Baier}, \citenamefont {Aikawa}, \citenamefont {Lepers},
  \citenamefont {Wyart}, \citenamefont {Dulieu},\ and\ \citenamefont
  {Ferlaino}}]{Becher2018pra}%
  \BibitemOpen
  \bibfield  {author} {\bibinfo {author} {\bibfnamefont {J.~H.}\ \bibnamefont
  {Becher}}, \bibinfo {author} {\bibfnamefont {S.}~\bibnamefont {Baier}},
  \bibinfo {author} {\bibfnamefont {K.}~\bibnamefont {Aikawa}}, \bibinfo
  {author} {\bibfnamefont {M.}~\bibnamefont {Lepers}}, \bibinfo {author}
  {\bibfnamefont {J.-F.}\ \bibnamefont {Wyart}}, \bibinfo {author}
  {\bibfnamefont {O.}~\bibnamefont {Dulieu}}, \ and\ \bibinfo {author}
  {\bibfnamefont {F.}~\bibnamefont {Ferlaino}},\ }\href {\doibase
  10.1103/PhysRevA.97.012509} {\bibfield  {journal} {\bibinfo  {journal} {Phys.
  Rev. A}\ }\textbf {\bibinfo {volume} {97}},\ \bibinfo {pages} {012509}
  (\bibinfo {year} {2018})}\BibitemShut {NoStop}%
\bibitem [{\citenamefont {Clark}\ and\ \citenamefont
  {Greene}(1999)}]{ClarkGreene1999rmp}%
  \BibitemOpen
  \bibfield  {author} {\bibinfo {author} {\bibfnamefont {W.}~\bibnamefont
  {Clark}}\ and\ \bibinfo {author} {\bibfnamefont {C.~H.}\ \bibnamefont
  {Greene}},\ }\href@noop {} {\bibfield  {journal} {\bibinfo  {journal} {Rev.
  Mod. Phys.}\ }\textbf {\bibinfo {volume} {71}},\ \bibinfo {pages} {821}
  (\bibinfo {year} {1999})}\BibitemShut {NoStop}%
\bibitem [{\citenamefont {Watanabe}\ and\ \citenamefont
  {Greene}(1980)}]{Watanabe1980pra}%
  \BibitemOpen
  \bibfield  {author} {\bibinfo {author} {\bibfnamefont {S.}~\bibnamefont
  {Watanabe}}\ and\ \bibinfo {author} {\bibfnamefont {C.~H.}\ \bibnamefont
  {Greene}},\ }\href {\doibase 10.1103/PhysRevA.22.158} {\bibfield  {journal}
  {\bibinfo  {journal} {Phys. Rev. A}\ }\textbf {\bibinfo {volume} {22}},\
  \bibinfo {pages} {158} (\bibinfo {year} {1980})}\BibitemShut {NoStop}%
\bibitem [{Top()}]{Toptica2021}%
  \BibitemOpen
  \href@noop {} {}\bibinfo {note} {Toptica Photonics AG, private
  communication}\BibitemShut {NoStop}%
\end{thebibliography}
\end{document}